\documentclass[a4paper,12pt]{article}
\usepackage[numbers,sort,compress]{natbib}

\usepackage[margin=1in]{geometry}

\usepackage[utf8]{inputenc}
\usepackage[T1]{fontenc}
\usepackage{amsmath}
\usepackage{amsfonts}
\usepackage{amssymb}
\usepackage{graphicx}
\usepackage{hyperref}
\usepackage{epstopdf}
\usepackage{tikz-feynman}
\usepackage{xcolor}
\usepackage{subcaption}
\usepackage{tikz-cd}
\usepackage{array, booktabs}

\usepackage[english]{babel}

\begin{document}

\title{\bf{Preons, Braid Topology, and Representations of Fundamental Particles}}

\author{David Chester$^{1, }$\footnote{\url{DavidC@QuantumGravityResearch.org}}  { }  Xerxes D.~Arsiwalla$^{2, }$\footnote{\url{x.d.arsiwalla@gmail.com}}  { }  Louis H.~Kauffman$^{3, 
}$\footnote{\url{loukau@gmail.com}}    \\  
{}  \\
{\it \small $^{1}$Quantum Gravity Research, Topanga, CA, USA}\\ 
{\it \small $^{2}$Wolfram Institute for Computational Foundations of Science, Champaign, IL, USA} \\    
{\it \small $^{3}$University of Illinois at Chicago, Chicago, IL, USA} 
}

\date{}

\maketitle

\begin{abstract}
In particle phenomenology, preon models study compositional rules of standard model interactions. In spite of empirical success, mathematical underpinnings of preon models in terms of group representation theory have not been fully worked out. Here, we address this issue while clarifying the relation between different preon models. In particular, we focus on two prominent models: Bilson-Thompson's helon model, and Lambek's 4-vector model. We determine the mapping between helon model particle states and representation theory of Lie algebras. Braided ribbon diagrams of the former represent on-shell states of spinors of the Lorentz group. Braids correspond to chirality, and twists, to charges. We note that this model captures only the $SU(3)_c\times U(1)_{em}$ sector of the standard model. We then map the twists of helon diagrams to the weight polytope of $SU(3)_c \times U(1)_{em}$. The braid structure maps to chiral states of fermions. We also show that Lambek's 4-vector can be recovered from helon diagrams. Alongside, we introduce a new 5-vector representation derived from the weight lattice. This representation contains both, the correct interactions found in 4-vectors and the inclusion of chirality found in helons. Additionally, we demonstrate topological analogues of  CPT transformations in helon diagrams. Interestingly, the braid diagrams of the helon model are the only ones that are self-consistent with CPT invariance. In contrast to field-theoretic approaches, the compositional character of preon models offers an analogous particle-centric perspective on fundamental interactions. 

\vspace{2pc}
{\it Keywords}: Preons, Standard Model Phenomenology, Braid Topology, Representation Theory.  
\end{abstract}

\clearpage

\tableofcontents

\section{Introduction}

Fundamental particles in standard quantum field theory are typically regarded as point-structures. Even though contemporary proposals of physics at the Planck scale have often toyed with the idea of some sort of internal structure for fundamental particles, based on the notion of a possible minimal length scale (see \cite{hossenfelder2013minimal} for a comprehensive overview), no conclusive evidence to this effect has been found to date \cite{kumar2020quantum}. In fact, following up on early ideas of A.~Lees in 1939 \cite{lees1939xxxvi}, it was P.A.M. Dirac himself, who in 1962 proposed a model for fermions as spheres extended in three spatial dimensions \cite{dirac1962extensible}. Part of Dirac's motivation also stemmed due to considerations of quantum gravity \cite{dirac1962particles}. Though Dirac conceded that this spherical particle model produced an incorrect value for the proposed mass of the muon \cite{dirac1962extensible}, the search for possible internal structure associated with fundamental particles continued. Earlier in 1931, it was Dirac again who proposed a one dimensional singularity in the electromagnetic field in order to explain charge quantization of the electron \cite{dirac1931quantised}. This singularity is what is known as the Dirac string, and it assumed the existence of magnetic monopoles. Of course, the Dirac string was not intended as an observable "extended particle", but is understood as a gauge dependent topological construct reflecting the non-simply connectedness of the $U(1)$ gauge field, resulting in the vector potential being well-defined only patch-wise. 

Later, several particle physicists in the 1970's and 80's proposed and studied "preon" models \cite{Pati:1974yy,Shupe:1979fv,Harari:1979gi,Harari:1980ez,Raitio:1979ru,Lehto:1981uz,Fritzsch:1981zh,Bars:1982zq,Zenczykowski:2008xt,Raitio:2018ofm}. A preon was thought of as an internal structure or building block of a particle, whose composites  realized the observable particles. The earliest realizations of preons  were discussed in the works of Pati and Salam (1974), Shupe (1979), Harari (1979), and Raitio (1980)    \cite{Pati:1974yy,Shupe:1979fv,Harari:1979gi,Harari:1980ez,Raitio:1979ru}. Harari called these structures "rishons" and formulated their model based on two binary symbols indicating one-third of a positive charge, and one-third of a negative charge \cite{Harari:1979gi}.  In 2000, J. Lambek formalized the Harari's rishon model in terms of a 4-vector representation  \cite{Lambek:2000ek}. Here, interactions between particles were realized as vector addition in an abstract four-dimensional space \cite{Lambek:2000ek}. Remarkably, Lambek's 4-vector model correctly reproduced part of the standard model interactions involving first generation fermions. The main drawback with this model (besides applying to only first-generation fermions) was that it was agnostic to chirality, that is, it did not distinguish between left and right-handed fermions. 

In 2005, Bilson-Thompson proposed a new representation of fundamental particles based on 3-stranded ribbon braids with chiral fermionic states \cite{Bilson-Thompson:2005mby}, which was further explored in subsequent works \cite{Bilson-Thompson:2006xhz,Bilson-Thompson:2008cmf,Bilson-Thompson:2009dvo,Bilson-Thompson:2011hue,Bilson-Thompson:2012bvs}. Interestingly, here topology plays an important role in distinguishing particle chirality. This model was called the "helon model", where individual strands or ribbons of the braided representation were referred to as helons. These ribbons could cross each other as particular ways, as well as twist in itself. The crossings distinguish chirality, whereas, the twisting encodes electric charge. 

While attempting to provide a systematic synthesis of phenomenologically verifiable features of preon models, in this work, we show how these models relate to representation theory; thereby providing an understanding of why they might have worked to reproduce some of the known particle interactions of the standard model. 
In addition, we demonstrate that the braid topology as a representation of particle chirality and particle/antiparticle identification can be understood in terms of the mapping from the braid group $\mathcal{B}_3$ to $SL (2, \mathbb{Z})$, which in turn sits within $SL (2, \mathbb{C})$, the double cover of the restricted Lorentz group. The values from the twist and braid operators are found from linear transformations of the weight-lattice coordinates associated with the representation theory of particles with $SL(2,\mathbb{C})\times SU(3) \times U(1)$ symmetry. 

Now, the braid diagrams of the helon model remarkably reproduce part of the standard model interactions involving first generation fermions, while also taking into account chirality of particles. However, a number of issues still persist. The relationship of the helon model to representation theory of groups had never been articulated before. In this work, we clarify this relation in terms of weight lattices of Lie algebras. Our analysis suggests that the helon model (as well as the 4-vector model) only captures the $SU(3)_c \times U(1)_{em}$ interactions, rather than the whole standard model. Moreover, up until now, the precise relation between the 4-vector model and the helon model remained unclear due to a slight mismatch of certain particle assignments. Here again, we shall demonstrate that a specific modification of the original helon model rectifies this mismatch. Furthermore, we also clarify how C, P and T transformations are realized in the helon model. One may ask why certain braid diagrams give a representation of fundamental particles? To address this, we show that the braid diagrams of the helon model are the only ones that are self-consistent with CPT invariance, which provides a natural motivation for such a braid representation. 

In terms of its broader significance, the relevance of preon models rests on the fact that these are purely compositional process theories (or models) with no additional spatiotemporal, mechanistic or dynamical structure; only combinatorial rules governing how parts of the system interact with each other to compose together as processes. Compositional theories are often formulated as diagrammatic calculi  based on the underlying process algebra. In fact, compositional process-theoretic approaches have been extensively used in applications ranging from quantum circuit analysis \cite{coecke2011interacting,gorard2020zx,gorard2021zx}; tensor network analysis, including representation of higher-order processes \cite{montangero2018introduction,zapata2022invitation,zapata2023hypermatrix,zapata2024diagrammatic}; and also category-theoretic approaches relating to spacetime causality \cite{arsiwalla2020homotopic,arsiwalla2021homotopies,arsiwalla2023pregeometry,rickles2023ruliology,arsiwalla2024pregeometric}. At their core, such approaches eschew explicit geometry in favor of algebraic or topological structures  \cite{arsiwalla2024operator,chester2024quantization}, which is also the way one constructs prospective topological quantum field theories discussed in high-energy physics and computes observables using compositional rules   \cite{carter1999structures,aganagic2005topological,arsiwalla2006phase,arsiwalla2008more,arsiwalla2009entropy,arsiwallasupersymmetric}. In the current context of particle physics, the compositionality inherent in preon models provide a packaging of combinatorial properties governing fundamental interactions. Moreover, as we shall show, these combinatorial rules are in fact related to the underlying gauge symmetry and the associated group representation theory. Compared to the standard field-theoretic perspective, the combinatorics of preon models provide an analogous particle-centric perspective on standard model interactions based on symmetry groups and their weight lattices.   

The outline of this paper is as follows: Section \ref{helon} introduces the helon model. Section \ref{Lambek} introduces Lambek's 4-vector model and compares it with the helon model. Section \ref{CPT} proposes candidates for $C$, $P$, and $T$ transformations in the helon model and demonstrates that $CPT$ symmetry uniquely identifies the fermionic braid states. Section \ref{interactions} demonstrates that the helon model successfully contains the gauge bosons of $SU(3)_c\times U(1)_{em}$. Section \ref{weight-lattice} establishes the relationship between representation theory in the weight lattice to the helon model and Lambek's model. Section \ref{conc} ends with our concluding remarks.

\section{The Helon Model}
\label{helon}

\begin{figure}[h!]
\centering
\includegraphics[width=0.8\linewidth]{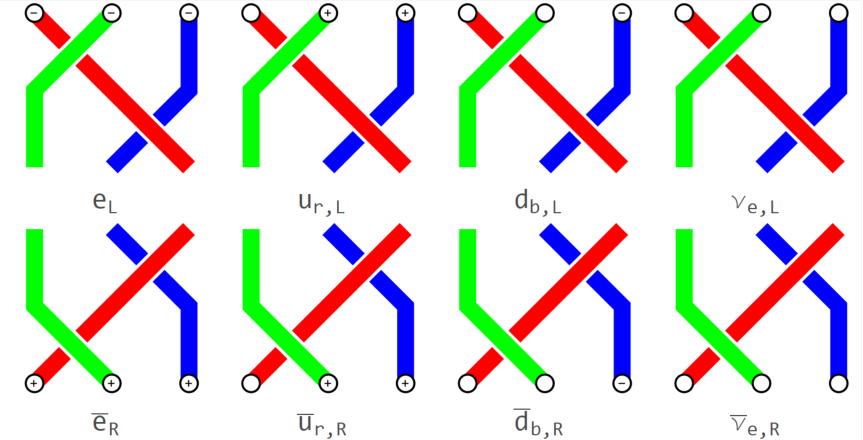}
\includegraphics[width=0.7\linewidth]{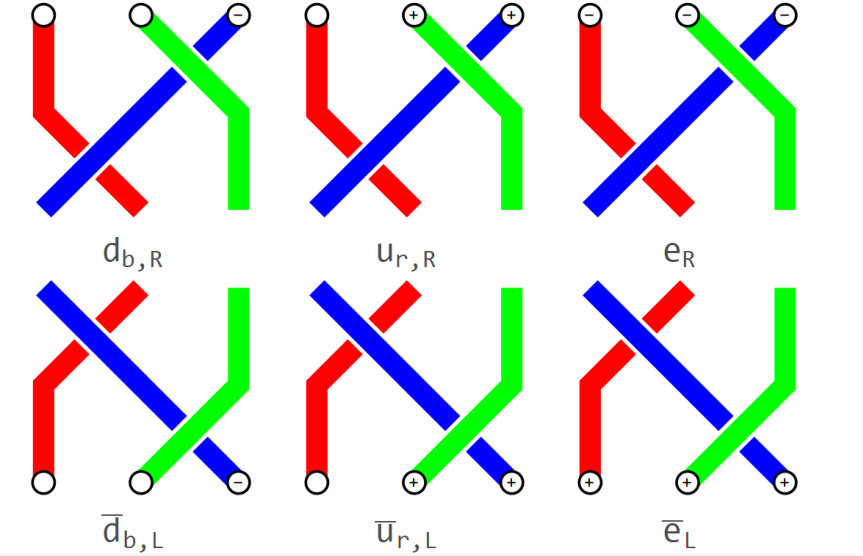}
\caption{Fermion assignments for a revised helon model are shown above. The up-quark states are updated to ensure the correct interactions. }
\label{updated-braids}
\end{figure}

The helon model depicts elementary particles as a braid diagram of three twisted ribbons. The spinor degrees of freedom such as chirality and particle/anti-particle are encoded by the braids, while the charges are encoded by the twists of the ribbons. CPT invariance of particle states in this model will involve both, the braid data, and the twists in helon model diagrams. The original paper presented four different braid diagrams with two braid operators \cite{Bilson-Thompson:2005mby}, while later work appeared to present only two \cite{Bilson-Thompson:2006xhz}. 

\subsection{Particle Representation with Helons }

The fermion assignments of our revised helon model are shown in Fig.~\ref{updated-braids} (the revisions we have proposed here will all be justified in the next few sections).  
The electron/positron states include both,  a left- and right-chiral electron/positron; that is, four on-shell states in total, which are encoded by four unique braid states. The braid group $\mathcal{B}_3$ of three ribbons contains generators $\sigma_1$ as braiding the 1st strand over the 2nd, $\sigma_2$ as the 2nd over the 3rd, $\sigma_1^{-1}$ as the 1st under the 2nd, and $\sigma_2^{-1}$ as the 2nd under the 3rd. The charges are uniquely determined by the twist configurations on each ribbon. 

In terms of braid generators, $\sigma_2 \sigma_1^{-1}$ is the left-chiral particle and $\sigma_1 \sigma_2^{-1}$ is the right-chiral anti-particle, which form two on-shell degrees of freedom of a Weyl spinor. The generators $\sigma_1^{-1}\sigma_2$ gives the right-chiral particle, while $\sigma_2^{-1} \sigma_1$ provides the left-chiral anti-particle. This assignment agrees with Bilson-Thompson except for the up-quark family. We will later demonstrate that our revised particle assignment indeed agrees with standard model phenomenology. 

The first generation of standard model fermions were presented in the original Bilson-Thompson model. Each ribbon may be twisted in a clockwise or counterclockwise manner once, yet the braids never contain oppositely twisited ribbons. This uniquely identifies fifteen different charges, such as $(000)$, $(00\pm)$, $(0\pm 0)$, $(\pm 00)$, $(0\pm\pm)$, $(\pm 0\pm)$, $(\pm\pm 0)$,  and $(\pm\pm\pm)$, which is the number of unique fermionic charges in the standard model or alternatively, $SU(5)$ GUT (with no right-handed neutrinos). Bilson-Thompson could have included the right-handed neutrinos just as well, but those have been  left out. The four valid braid states include the oppositely charged states, that is, 30 on-shell states are identified for  first generation standard model particles  in the helon model. 

\begin{figure}
\centering
\includegraphics[width=0.7\linewidth]{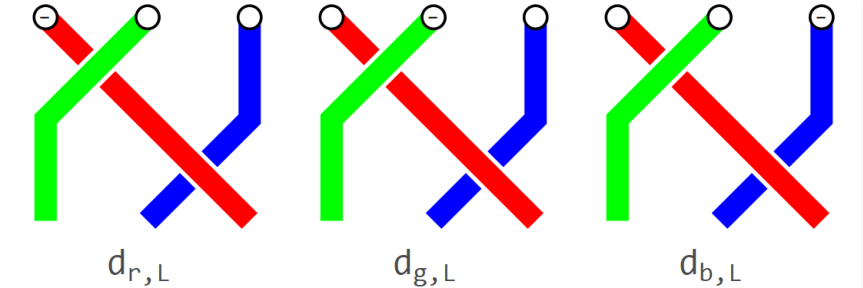}
\includegraphics[width=0.7\linewidth]{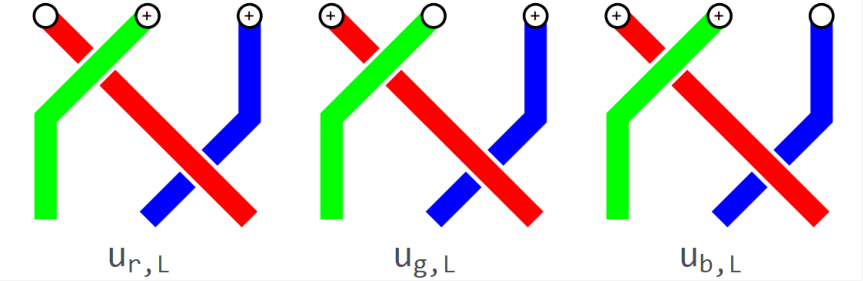}
\caption{The three color states of the left-chiral down and up quarks are shown above.}
\label{down-braids}
\end{figure}

\begin{figure}[h!]
\centering
\begin{subfigure}[t]{0.25\linewidth}
\includegraphics[width=\linewidth]{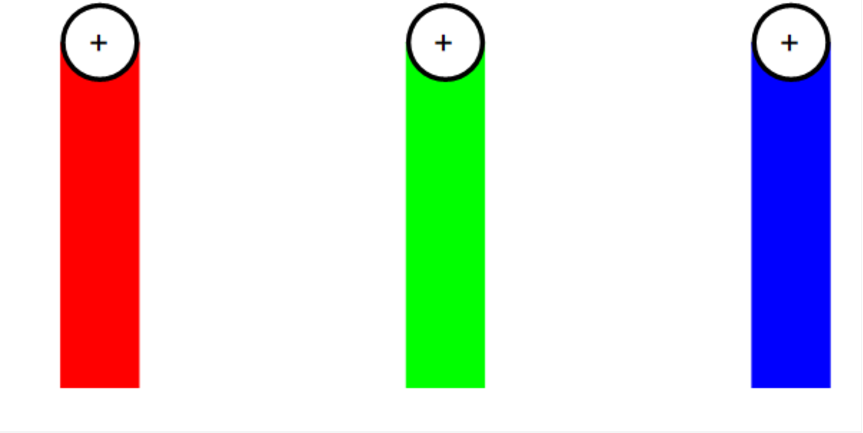}
\caption{$W^+$ boson}
\end{subfigure}
\qquad\qquad
\begin{subfigure}[t]{0.25\linewidth}
\includegraphics[width=\linewidth]{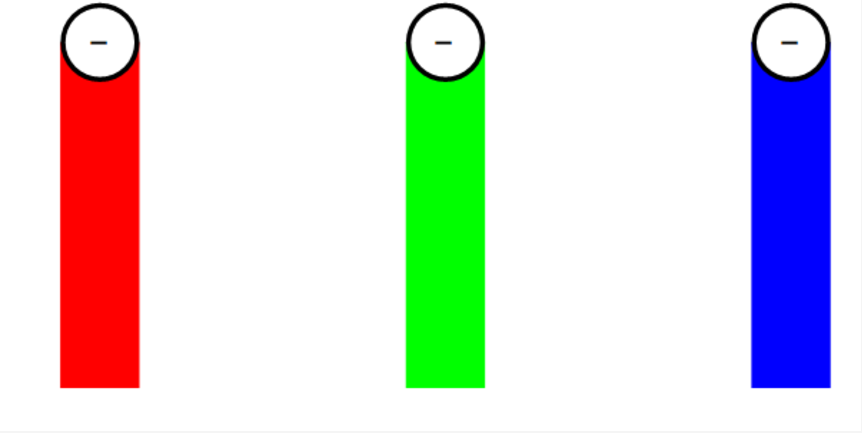}
\caption{$W^-$ boson}
\end{subfigure}
\\
\begin{subfigure}[b]{0.25\linewidth}
\includegraphics[width=\linewidth]{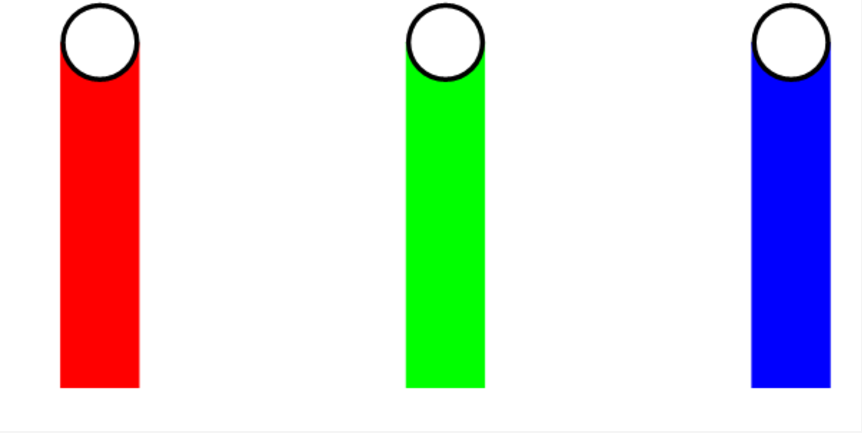}
\caption{Photon ($\gamma$)}
\end{subfigure}
\qquad\qquad
\begin{subfigure}[b]{0.25\linewidth}
\includegraphics[width=\linewidth]{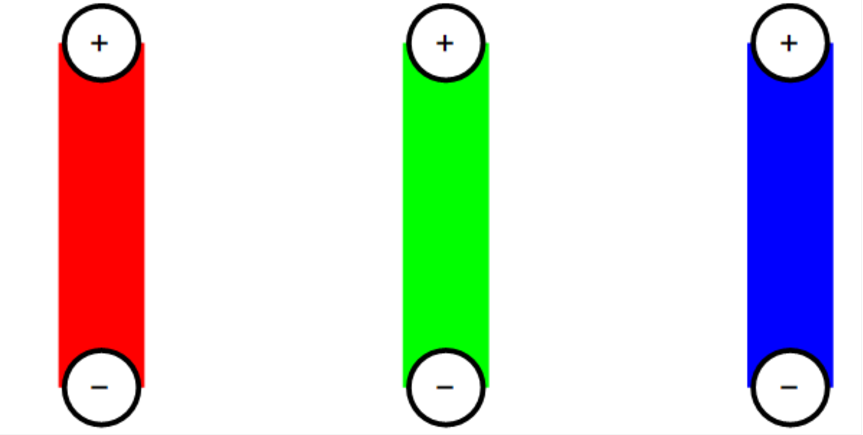}
\caption{$Z$ boson}
\end{subfigure}
\caption{The braids claimed to be associated with the electroweak sector below the electroweak scale. Note that the Bilson-Thompson model does not have any data to associate the weak force with left-chiral fermions. }
\label{electroweak-braids}
\end{figure}

The gauge bosons are also assigned braided ribbon structure, but the gauge bosons are only given twists and are not associated with any braid operators. The electric charge of the standard model gauge bosons was used to determine the assignments of the twist. The primary difference between the photon and the Z-boson is that the Z-boson is a mixture of the $W^0$ and $B$ bosons such that it acquires mass and couples to left- vs right-chiral fermions differently. To capture this difference, Bilson-Thompson proposes that the electrically-neutral Z-boson combines a twist/anti-twist that does not annihilate each other, yet is found on each of the three ribbons without any braiding.  The proposed $W^\pm$ bosons in the helon model interact with both left- and right-chiral fermions, which is problematic. Fig.~\ref{electroweak-braids} summarizes the electroweak gauge bosons in the helon model. Our analysis demonstrates that the Bilson-Thomspon model contains the standard model fermionic sector, but only captures the $SU(3)_c\times U(1)_{em}$ gauge bosons, as there is nothing preventing right-chiral weak interactions and the $Z$-boson is topologically equivalent to the photon.

\begin{figure}[h!]
\centering
\includegraphics[width=\linewidth]{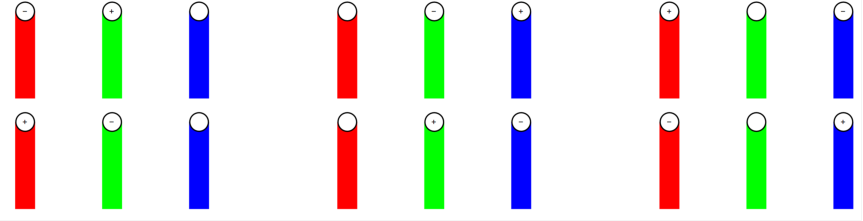}
\caption{Braid diagrams for gluons corresponding to the six non-Cartan generators of $SU(3)$.}
\label{strong-braids}
\end{figure}

The gluons are found in a manner similar to the Z-boson, except a mix of positive, negative, and zero twist are combined to give electrically neutral states for the six non-Cartan generators. The two Cartan generators of $SU(3)$ found as diagonal $3\times 3$ matrices in the Gell-Mann basis. Since the Cartan generators involve linear combinations of different color/anti-color pairs to remain as traceless generators of $SU(3)$, it is not clear if a single braid diagram can represent these generators. Similarly, there was no braid diagram for the $W^0$ boson corresponding to a Cartan generator, but this was more due to the desire for identifying $\gamma$ and $Z$ bosons below the electroweak scale. The braid diagrams for the eight gluons in the helon model are shown in Fig.~\ref{strong-braids}. The braids shown for the Cartan generators are merely suggestive; a more accurate depiction may require treating the Bilson-Thompson braids as basis elements in some ring or algebra to allow for linear combinations.

\subsection{Composition Laws with Helons }

While discussing compatibility between the helon and 4-vector models, there is one important issue that  has to be addressed; namely, the nature of their respective composition laws. While the 4-vector model assumes a fully commutative composition rule by way of vector addition; the helon model is noncommutative, that is, the order of braids composing with each other does matters. In order to understand how these two differing models may  still corroborate to reproduce correct standard model interactions, we need to closely examine their composition rules. 

Now, in the 4-vector model, the composition rule for particle states being ordinary vector addition (the 4-vector model is discussed in detail in the following section), is commutative. However, helon diagrams, being a set of three braided ribbons with twists, is in general noncommutative. A helon model diagram can be expressed using the following notation: $\vec{t} \, \sigma$, where $\vec{t}$ denotes the 3-tuple of twists on the three respective ribbons of the diagram, and $\sigma$ denotes the braid word. With this, particle composition in the helon model is obtained via multiplication of two helon diagrams as follows
\begin{eqnarray}
    ( \vec{t} \, \sigma ) \cdot  ( \vec{t}^{\prime} \, \sigma^{\prime} ) = ( \vec{t} \bigoplus \vec{t}^{\prime \, \sigma} ) \, \sigma \, \sigma^{\prime}
\end{eqnarray}
where $\vec{t}^{\prime \, \sigma}$ denotes a permutation of the components of $\vec{t}^{\prime}$ by the action of the braid $\sigma$; the  $\bigoplus$ denotes standard vector addition, and $\sigma \, \sigma^{\prime}$ yield braid multiplication placing $\sigma^{\prime}$ below $\sigma$. Such a multiplication is a purely topological operation of braided ribbons with twists. 

Now, at first glance, it might appear that the noncommutativity of  braid operators disallows for consistent composites of particles to be constructed from  multiplication of helon model diagrams. This is because both color and electric charges are determined by the twists. Since hadrons are colorless, each ribbon should contain the same total number of twists. Taking the particle assignments from Fig.~\ref{updated-braids} and combining quarks with braid multiplication in different orders can lead to different numbers of twists on each strand. Combining quark braids to build hadrons with braid multiplication in this way leads to some permutations of quarks to work, while others do not.

\begin{figure}[h!]
\centering
\begin{subfigure}{0.9\linewidth}
    \centering
    \includegraphics[width=0.45\linewidth]{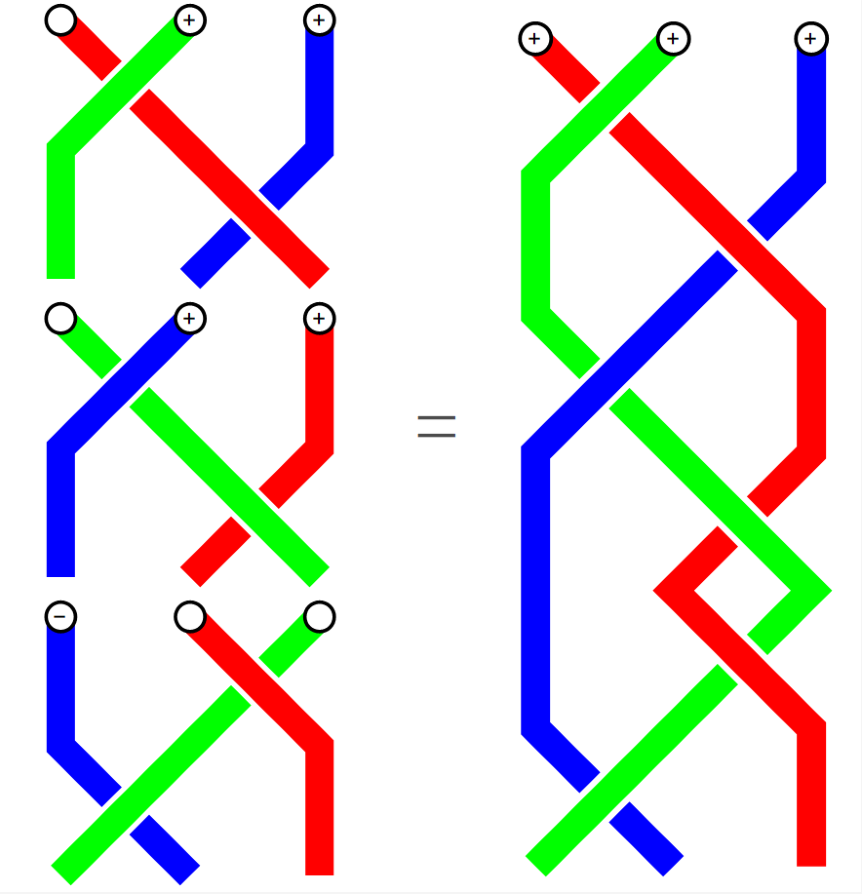}
    \hfill
    \includegraphics[width=0.45\linewidth]{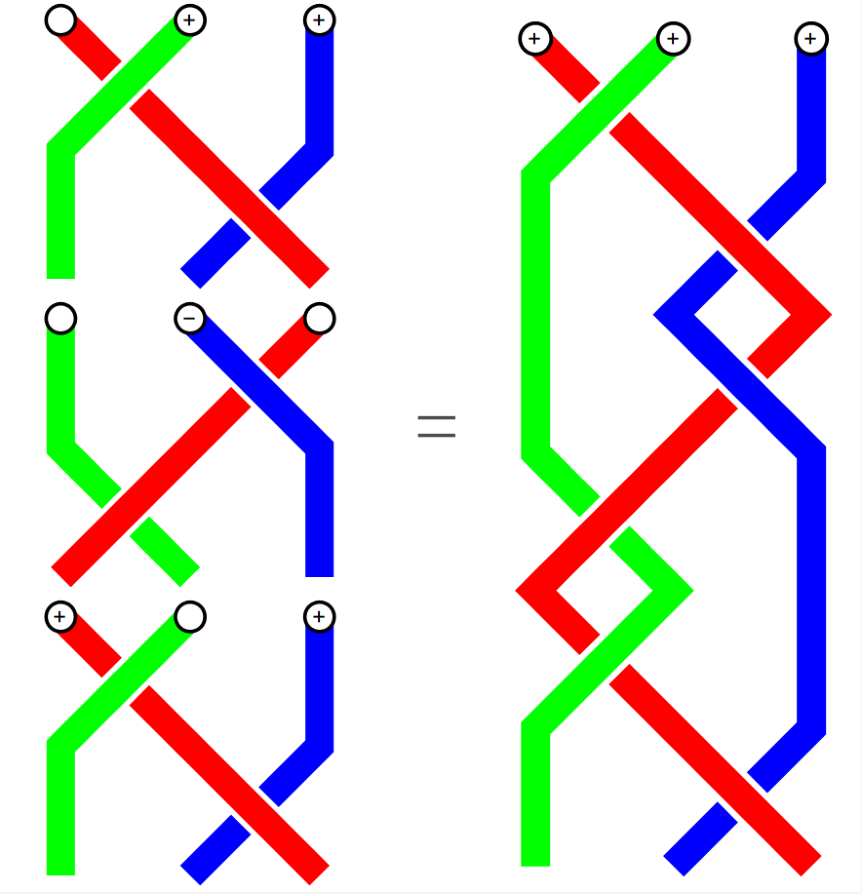}
    \subcaption{Two examples of proton braids are shown above. On the left, $u_{L,r}u_{L,g}d_{R,b}$ is depicted (from the top to the bottom). On the right, $u_{L,r}d_{R,b}u_{L,g}$ is shown. Permuting $u_{L,g}$ and $d_{R,b}$ leads to the same number of twists per ribbon.}
\end{subfigure}

\vspace{1em}

\begin{subfigure}{0.9\linewidth}
    \centering
    \includegraphics[width=0.45\linewidth]{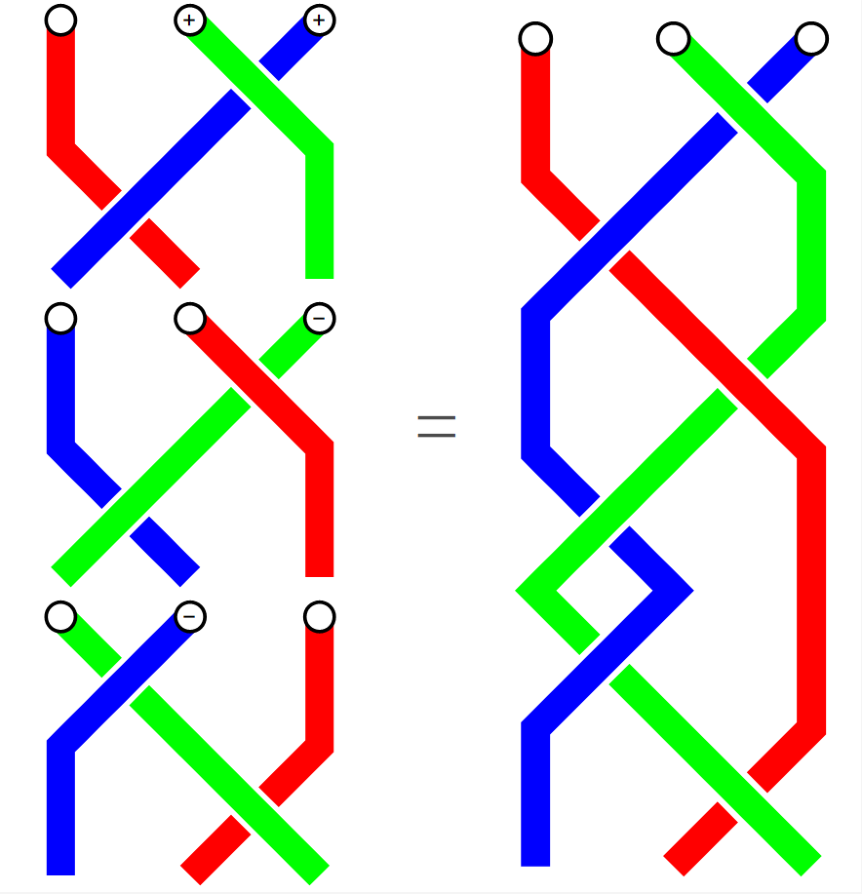}
    \hfill
    \includegraphics[width=0.45\linewidth]{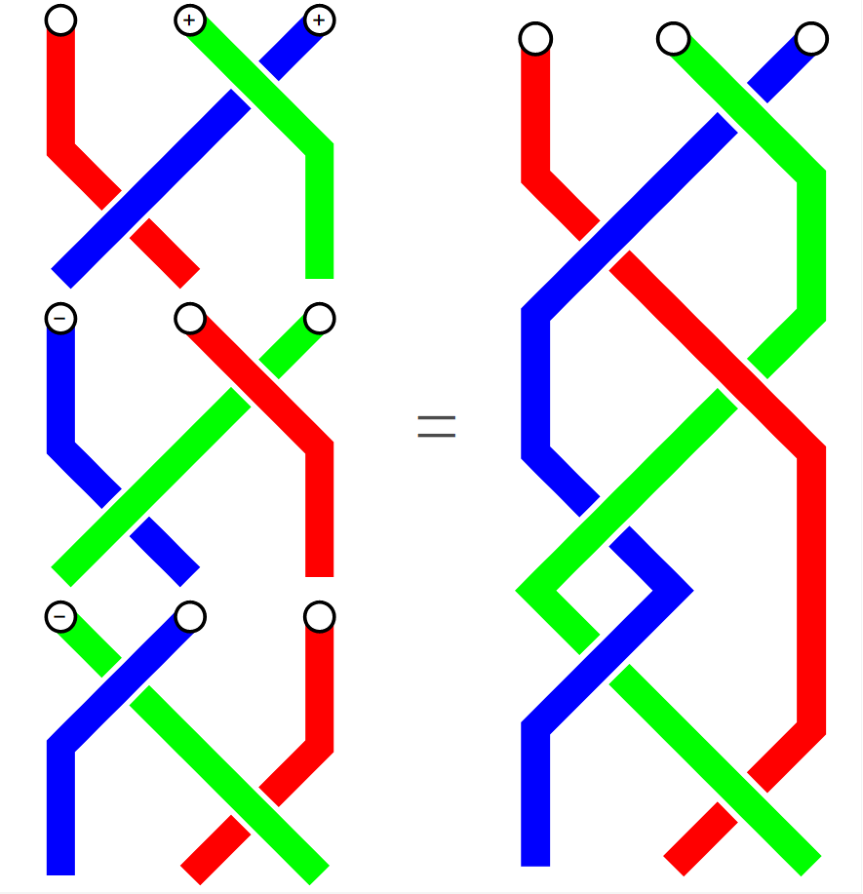}
    \subcaption{Two examples of neutrons are shown above. On the left, $u_{R,r}d_{R,g}d_{L,b}$ is shown, while $u_{R,r}d_{R,b}d_{L,g}$ is shown on the right. Here, only the color (not chirality) of the down quarks are permuted. Once again, both braids lead to the same number of twists per ribbon.}
\end{subfigure}

\caption{Candidates for quark composite braids are shown above. If color-matched braid multiplication was not used, then the proton braids would not always have one twist per ribbon and the neutron braids sometimes could contain twisted ribbons.}
\label{fig:proton_neutron_braids}
\end{figure}

Remarkably, it turns out that the above issue (for consistently constructing hadron states) is completely resolved by additionally declaring (compared to the original Bilson-Thompson model) that the color of a particle is not fixed to the first, second, or third strand of each braid. Rather, we first freely assign red, green, and blue to the first  braid diagram in the composite;  subsequently, the second braid diagram gets the color of the ribbons assigned from the permutation from the first particle's braid. The third particle's braid then gets color from the permutation of the second particle's braid, and so on. In other words, entire strands, in composite braid diagrams, maintain the same color. We will refer to this multiplication as "color-matched braid multiplication".  As a concrete example,  Fig.~\ref{fig:proton_neutron_braids} shows how a proton and neutron braid can be represented by combining various combinations of quark braids of different color, chirality, and braid permutation. We confirmed that all possible combinations of quarks that lead to a colorless hadron indeed lead to the same number of twists on each ribbon. 

If braid multiplication is done without the above color matching, one obtains physically incorrect composites such as an apparently colorless state from three quarks that are all red. This further demonstrates that if braid multiplication is used to model composite particles states in the helon model, then color-matched braid multiplication should be used. Note that permuting particle braids in a composite made from color-matched braid multiplication will also produce different topological braids. Thus a proton will, in this model, have more than one possible topology. However, the twists of the ribbons will be identical for all proton braids. 

Bilson-Thompson composites were represented more abstractly as a direct sum of braids, but this choice was not shown to be satisfactory. As we will show next, Lambek obtains a model for composites that is identical to how he obtains interactions. This makes intuitive sense, as the only difference between a particle decay and a bound state is whether the particles are free or bound to each other. As such, we suggest a modification to the original helon model with composites treated using color-matched braid multiplication rather than a direct sum or addition. By using color-matched braid multiplication for composites in the helon model, the twists on the ribbons are found to add just like they do in Lambek's model.

\section{The 4-Vector Representation}
\label{Lambek}

Prior to the helon model, in the year 2000, Joachim Lambek had introduced a 4-vector representation of fundamental particles \cite{Lambek:2000ek}. Lambek's model was intended as a formalization of the earlier rishon model proposed by Harari (which was also conceptually similar to Shupe's and Raitio's proposals)  \cite{Shupe:1979fv,Harari:1979gi,Raitio:1979ru}. In Lambek's 4-vector representation, shown in Table \ref{table:Lambek}, each lepton, quark, and gauge boson are associated with a fundamental 4-vector with entries $\pm 1$ or $0$. The first component of this 4-vector corresponds to the fermion number, while the remaining three sum up to the electric charge and provide a 3-vector for color. A particle interaction, that is, a vertex of a Feynman diagram, is realized via the addition of two 4-vectors. In this sense, interaction processes in this model are completely commutative.  However, the 4-vector model does not account for particle chirality, nor does it accommodate fermions beyond the first generation of the standard model. On the other hand, the helon model does account for chirality via its braided structure. Nonetheless, to the credit of the 4-vector model, it reproduced many of the first-generation standard-model interactions, albeit whilst ignoring particle chirality in those processes.

\begin{table}[h!]
\centering
\begin{tabular}{l |l |l| l}
\toprule
\textbf{Fermions} & \textbf{Antifermions} & \textbf{Gauge bosons} & \textbf{Baryons} \\
\midrule
$e^-: 1 - i - j - k$ & $\bar{e}^+: -1 + i + j + k$ & $\gamma, Z^0: 0$ & $n: 3$ \\[6pt]
$\nu: 1$ & $\bar{\nu}: -1$ & $W^-: -i - j - k$ & $p^+: 3 + i + j + k$ \\[6pt]
$u_r: 1 + j + k$ & $\bar{u}_r: -1 - j - k$ & $W^+: i + j + k$ &  \\[6pt]
$u_b: 1 + i + k$ & $\bar{u}_b: -1 - i - k$ & $G_{\bar{b}g}: j - k$ &  \\[6pt]
$u_g: 1 + i + j$ & $\bar{u}_g: -1 - i - j$ & $G_{\bar{r}b}: -i + k$ &  \\[6pt]
$d_r: 1 - i$ & $\bar{d}_r: -1 + i$ & $G_{\bar{b}r}: i - j$ &  \\[6pt]
$d_b: 1 - j$ & $\bar{d}_b: -1 + j$ & $G_{\bar{g}b}: -j + k$ &  \\[6pt]
$d_g: 1 - k$ & $\bar{d}_g: -1 + k$ & $G_{\bar{b}r}: i - k$ &  \\[6pt]
 &  & $G_{\bar{r}g}: -i + j$ &  \\
\bottomrule
\end{tabular}
\caption{Fermions, antifermions, gauge bosons, and baryons with Lambek’s quaternionic (4-vector)   assignments.}
\label{table:Lambek}
\end{table}

An interesting feature about the 4-vector model is that it even accounts for composite states of first-generation fundamental particles using the same additive composition law. For instance, it neatly accounts for neutron decay. For instance, the 4-vector corresponding to the neutron $n$ and proton $p^+$ as quark composites yields $n = u_r + d_g + d_b$ and $p^+ = u_r + u_g + d_b$ respectively (any other permutation of colors works just as well). Substituting the 4-vectors for the above up and down quarks from Table \ref{table:Lambek} yields $n = 3$ and $p^+ = 3 + i + j + k$ respectively. Next, using the 4-vectors for the electron $e^-$ and its anti-neutrino $\overline{\nu}$ from Table \ref{table:Lambek}, one sees that the equation $n = p^+ + e^- + \overline{\nu}_e$ is indeed satisfied as a 4-vector equation.

\subsection{Mapping Helons to Four-Vectors }

Now, the helon model with its braided structure is certainly not a vector model. In the former,  interactions are realized via color-matched braid multiplication. The twists correspond to electric charges, while the braid crossings capture chirality. Even though the helon model carries some additional data than the 4-vector model, it is useful to highlight the compatibility between the two models. 

Before we elaborate on the precise mapping between the models, let us first draw attention to the revision (compared to the original particle assignments by Bilson-Thompson) that we have  proposed for helon model particle assignments (already included in in Fig.~\ref{updated-braids} above). This revision turns out to be crucial for correctly getting interaction vertices involving the up and down quark. The 4-vector model when compared to the original prescription of Bilson-Thompson, in terms of mapping twists of helons to the last three components of the 4-vector, suggests this modification to Bilson-Thomson's original scheme: The braid diagrams corresponding to the up and anti-up quark ought to be flipped (also inverting the charges). This new assignment, which we have incorporated in this paper in Fig.~\ref{updated-braids}, perfectly corroborates the helon model with Lambek's 4-vector model, while maintaining the additional advantage of including chirality. Furthermore, with this new assignment, the helon model, like the 4-vector model, also reproduces the first generation standard model interactions corresponding to the electromagnetic and strong force (both these models only partially capture aspects of the weak interactions). With this assignment, the position of twists in helon diagrams along with the signs, exactly match the last three components of Lambek's 4-vector.

Now, let us show how the fermion number of the 4-vector model is also captured by braids of the helon model. In the 4-vector model the fermion number $F$ was assigned values $+1$ for particles, $-1$ for antiparticles and $0$ for bosons. The braids of the helon model capture this assignment diagrammatically as follows. First note that all helon model diagrams are composed from the following four braid monomials in $\mathcal{B}_3$: $\sigma_1$, $\sigma_2$, $\sigma_1^{-1}$ and $\sigma_2^{-1}$. Using these, one obtains a total of 16 distinct braid words of length 2. However, of these 16, only 4 appear as particle states in the helon model  (in a following section, we show that these 4 choices of helon braids are the only ones that happen to be invariant under a topological realization of CPT transformations on $\mathcal{B}_3$). These 4 helon diagrams take a rather specific form, and can be expressed as $\sigma_i^{a} \sigma_j^{b}$, where $a$ takes values in $\{ 1, -1\}$, and $b = -a$. Then the following expression yields Lambek's fermion number $F$:
\begin{eqnarray}
    F = a \cdot i + b \cdot j
\label{fnum} 
\end{eqnarray}
Given that  helon braids are alternating, the pair $(i , j)$ can only be $(1 , 2)$ or $(2 , 1)$. With this it is straightforward to see that the above expression for $F$ only yields $+1$, $-1$ or $0$ respectively  for fermion, anti-fermion and boson diagrams in the helon model. 

Furthermore, let us now show how the above topological expression for the fermion number can generalized to account for, not just fundamental particle states, but also composites such as protons and neutrons.  Let us now consider the most general braid word in $\mathcal{B}_3$ with $n$ crossings $\sigma_{i_1}^{Sg( \sigma_{i_1} )}  \cdots \sigma_{i_n}^{Sg( \sigma_{i_n} )}$, where $Sg( \sigma_{i_j} )$ refers to the signature of the braid generator at strand ${i_j}$. This signature is $+1$ for 
upper left to lower right over crossings, and $-1$ for 
upper left to lower right under crossings, where “upper” and “lower” refer to height on the page. Then, the general expression for the fermion number $F$ corresponding to compositions of helon model particle states can be obtained from the formula
\begin{equation}
F  = \sum_{j = 1}^n   Sg( \sigma_{i_j} )  \cdot {i_j}
\label{fnum2} 
\end{equation}
for the general braid word $\sigma_{i_1}^{Sg( \sigma_{i_1} )}  \cdots \sigma_{i_n}^{Sg( \sigma_{i_n} )}$ taken as a generic composition of multiple length 2 words.   

With the above equation, it works out that a composition of helon model diagrams  corresponding to 1 up and 2 down quarks, obtained by placing  diagrams one below the other, using color-matched braid multiplication, results in $F = 3$. Likewise, for the proton, composing 2 up and 1 down quark  diagrams results in $F = 3$.  And also, it is straightforwardly to see that analogous compositions with anti-quarks result in  $F = -3$ for both, the anti-neutron and anti-proton. 

With the above dictionary between helon model diagrams and 4-vectors, we obtain the following mapping from helons to 4-vectors. Consider a helon model diagram ${\cal D}$. This consists of the following data: a length 2 word in $\mathcal{B}_3$, $\sigma_i^{a} \sigma_j^{b}$ (using the notation above eq. (\ref{fnum}); and twists on the three ribbons, which we denote via the tuple $( t_1, t_2, t_3 )$, and which take values $+1$, $-1$ or $0$.  Then the map $\tau : \textrm{Helons} \to \textrm{4-vectors}$ is given by
\begin{eqnarray}
    \tau ( {\cal D} ) = F + t_1 \cdot i + t_2 \cdot j + t_3 \cdot k
\end{eqnarray}
where $F$ is given by eq. (\ref{fnum}). This mapping from helons to 4-vectors also works for generic composite particle states, where again, composite particles in the helon model are obtained via color-matched braid multiplication as described in an earlier  section. In this case, the fermion number $F$ is obtained via the general formula in eq. (\ref{fnum2}) above.

\section{CPT Invariance and Braid Topology }
\label{CPT}

The helon model only uses a subset of all possible twisted braid diagrams. Understanding a physical principle to choose these braids over another would provide further justification for the helon model. In this section, we review a set of topological transformations that seem to map to the C, P, and T transformations. While CPT transformations are found in the Lie group, these topological transformations are transformations that act on the twisted braid states. We demonstrate that the braid diagrams of the helon model (taking into account the modification of helon assignments mentioned earlier) are the only braid diagrams that turn out to be invariant under the these topological analogues of CPT transformations.

To see this let us first let us understand how C, P and T transformations can be formulated within this model. For example, in case of the electron, one expects
\begin{eqnarray}
C  \, : \,  e_L^- \to  e_L^+  \\
P  \, : \,  e_L^- \to  e_R^-  \\
T  \, : \,  e_L^- \to  e_R^+  
\end{eqnarray}
In terms of the braided diagrams of the helon model, these transformations can be explicitly realized as
\begin{eqnarray}
C  \, : \,  \left( \sigma^{-1}  \otimes  \mathbf{1}  \right)  \left( \mathbf{1}  \otimes  \sigma   \right)  &\to&  \left( \sigma  \otimes  \mathbf{1}  \right)  \left( \mathbf{1}  \otimes  \sigma^{-1}   \right)  \\
P  \, : \,  \left( \sigma^{-1}  \otimes  \mathbf{1}  \right)  \left( \mathbf{1}  \otimes  \sigma   \right)   &\to&     \left( \mathbf{1}  \otimes  \sigma   \right)   \left( \sigma^{-1}  \otimes  \mathbf{1}  \right)     \\
T  \, : \,  \left( \sigma^{-1}  \otimes  \mathbf{1}  \right)  \left( \mathbf{1}  \otimes  \sigma   \right)   &\to&    \left( \mathbf{1}  \otimes  \sigma^{-1}   \right)    \left( \sigma    \otimes  \mathbf{1}  \right)     
\end{eqnarray}
where the tensor product $ \left( \sigma^{-1}  \otimes  \mathbf{1}  \right)  \left( \mathbf{1}  \otimes  \sigma   \right)$ denotes the braid word $\sigma^{-1}_1  \sigma_2$ in $\mathcal{B}_3$, referring to  $e_L^-$ in the helon model. 

Note that the helon model consists of words in the semi-direct product $B_3  \rtimes  B_2$, where elements of $B_2$ consist of twists that act on the braided ribbons in $\mathcal{B}_3$. These twists refer to electric charge, whereas the braids capture particle chirality.  Hence, one can read the CPT transformation given above even more generally in the following way: The charge conjugation operator $C$ acts on any word in $B_3  \rtimes  B_2$ via switching of the braid crossings from over to under or vice-versa, accompanied by reversing the orientation of the twists, corresponding to flipping the signs of  charges in the helon model. The parity operator $P$ acts on words in $B_3  \rtimes  B_2$ via a reflection across the vertical axis passing through the center of the braid diagram in the helon model. And, the time reversal operator $T$ acts on words in $B_3  \rtimes  B_2$ via a reflection across the horizontal axis passing through the center of the braid diagram,  accompanied by reversing the orientation of the twists, which reverse signs of charges. 

\begin{figure}
\centering
\begin{subfigure}[t]{\linewidth}
\includegraphics[width=\linewidth]{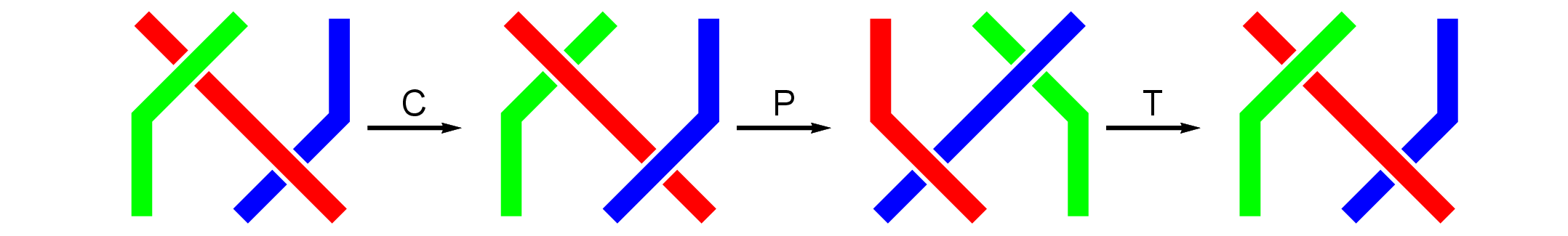}
\caption{The left-chiral braid state is invariant when applying the three operators above.}
\end{subfigure}
\\
\begin{subfigure}[t]{\linewidth}
\includegraphics[width=\linewidth]{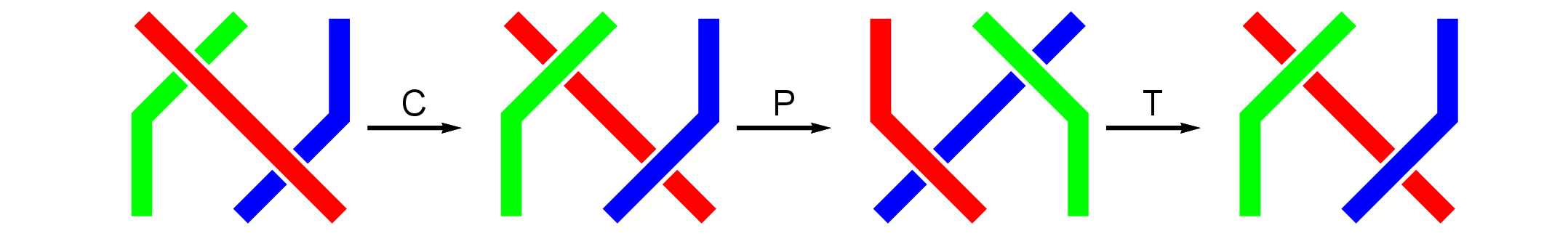}
\caption{A braid not associated with any particle is not invariant when applying the same three operators above.}
\end{subfigure}
\caption{The proposed CPT transformations leave invariant exactly those braids corresponding to helon model particle states. All other braid words are not invariant under these transformations.}
\label{CPTbraids}
\end{figure}

CPT invariance in the helon model can now be checked by applying the action of the $C$, $P$ and $T$ operators to braid words corresponding to particle states in the model. For instance, applying the proposed CPT transformations  upon the braid word corresponding to the $e_L^-$ in the helon model yields 
\begin{equation}
\begin{tikzcd}[column sep={3cm,between origins},row sep={3cm,between origins}]
\sigma^{-1}_1  \sigma_2   \rar["C"  ] &  \sigma_1  \sigma^{-1}_2   \rar["P"  ] &  \sigma^{-1}_2  \sigma_1   \rar["T"  ] &  \sigma^{-1}_1  \sigma_2  
\end{tikzcd} 
\end{equation}
corresponding to
\begin{equation}
\begin{tikzcd}[column sep={3cm,between origins},row sep={3cm,between origins}]
e_L^-   \rar["C"  ] &  e_L^+   \rar["P"  ] &  e_R^+   \rar["T"  ] &  e_L^- 
\end{tikzcd} 
\end{equation}
The braids associated with this sequence of mappings are shown in Fig.~\ref{CPTbraids}.

Similarly, these transformations work correctly for any other particle state specified within the helon model and for any order of application of $C$, $P$ and $T$ operators. Furthermore, what is interesting is that all other twisted braid words within $B_3  \rtimes  B_2$ which also consist of two crossings, but are not part of the helon model,  fail to be CPT invariant. For example, consider the braid word $\sigma_1  \sigma_2$, which is not included in the helon model. Applying CPT transformations to this gives 
\begin{equation}
\begin{tikzcd}[column sep={3cm,between origins},row sep={3cm,between origins}]
\sigma_1  \sigma_2   \rar["C"  ] &  \sigma^{-1}_1  \sigma^{-1}_2   \rar["P"  ] &  \sigma_2  \sigma_1   \rar["T"  ] &   \sigma^{-1}_1  \sigma^{-1}_2    
\end{tikzcd} 
\end{equation}
which is certainly not CPT invariant. The braid transformations for this are shown in Fig.~\ref{CPTbraids}. 

We have computationally checked that any other braid diagram not included in the helon model fails at being CPT invariant. There are 16 diagrams that include two braid operators on three strands. There are only four braid words that are invariant with respect to our proposed CPT transformation, which corresponds to $\sigma_1\sigma_2^{-1}$, $\sigma_2\sigma_1^{-1}$, $\sigma_1^{-1}\sigma_2$, and $\sigma_2^{-1}\sigma_1$.  
Thus, the particle states of the helon model, represented as braid diagrams with crossings in alternating positions and with opposite signatures are the only ones that happen to be invariant under the above-mentioned CPT  transformation.

Recall that the chirality of fermions is determined by braid group generators in $\mathcal{B}_3$. Now, conventional CPT transformations in QFT are known to be associated with $Pin(1,3)$, which is the double cover of $O(1,3)$. Interestingly, there exist a series of mappings that allows for $\mathcal{B}_3$ to be identified within $Pin(1,3)$. This can be seen starting from the homomorphism from $\mathcal{B}_3$ to the modular group $PSL(2,\mathbb{Z})$, where the generators of $\mathcal{B}_3$ are realized in $PSL(2,\mathbb{Z})$ via the identification  $\sigma_1 \to t$ and $\sigma_2 \to rtr$ (see \cite{salter2024ropes} for a recent overview). Here, $t = \begin{pmatrix}  1 & 1 \\  0 & 1  \end{pmatrix}$ and $r = \begin{pmatrix}  0 & -1 \\  1 & 0  \end{pmatrix}$ are generators of $PSL(2,\mathbb{Z})$. Note that $t$ and $rtr$ as matrices in $SL(2,\mathbb{Z})$ satisfy the braiding relations up to sign. Furthermore, $SL(2,\mathbb{Z})$ is included in $SL(2,\mathbb{R})$. That is in turn within $SL(2,\mathbb{C})$, and the latter within $Pin(1,3)$. Therefore, we have that 
\begin{equation}
B_3 \to{} SL(2,\mathbb{Z}) \hookrightarrow SL(2,\mathbb{R}) \hookrightarrow SL(2,\mathbb{C}) \hookrightarrow Pin(1,3)
\end{equation}
This provides indirect evidence that the proposed topological analogues of $C$, $P$, and $T$ transformations may correspond to the algebraic $C$, $P$, and $T$ transformations associated with $Pin(1,3)$.

\section{Standard Model Interactions in the Helon Model }
\label{interactions}

Next, the Feynman interaction rules of the standard model are exhaustively compared with the Bilson-Thompson model. We start by reviewing the gauge-boson interactions, followed by the fermions with the bosons.

\subsection{Gauge Boson Interactions}

The standard model contains non-Abelian gauge bosons that interact with $SU(3)\times SU(2)\times\times U(1)$ gauge symmetry. 
The Feynman rules for the strong- and weak-force gauge bosons stem from Yang-Mills theory. An arbitrary $SU(N)$ Yang-Mills gauge field $A_\mu^A$ corresponds to $G_\mu^A$ and $W_\mu^i$ for $N=2$ and $N=3$. 
The gluons and the W-bosons interact with themselves, respectively.
The B-boson sourced by weak hypercharge has no nonlinear self-interactions at the classical level to consider for interaction rules in Feynman diagrams. The Z-boson and photon are a mix of the $W^0$ and B-boson, which also interacts with $W^\pm$. 

\begin{figure}[h!]
\centering
\begin{subfigure}[t]{0.9\linewidth}
\centering
\begin{tikzpicture}
  \begin{feynman}
    \vertex (a);
    \vertex [above left=of a] (b) {\(G^{A}_{\mu}\)};
    \vertex [below left=of a] (c) {\(G^{C}_{\rho}\)};
    \vertex [right=of a] (d) {\(G^{B}_{\nu}\)};

    \diagram* {
      (b) -- [gluon] (a),
      (c) -- [gluon] (a),
      (a) -- [gluon] (d),
    };
  \end{feynman}
\end{tikzpicture}
\hspace{1cm} 
\begin{tikzpicture}
  \begin{feynman}
    \vertex (a);
    \vertex [above left=of a] (b) {\(G^{A}_{\mu}\)};
    \vertex [below left=of a] (c) {\(G^{D}_{\sigma}\)};
    \vertex [above right=of a] (d) {\(G^{B}_{\nu}\)};
    \vertex [below right=of a] (e) {\(G^{C}_{\rho}\)};

    \diagram* {
      (b) -- [gluon] (a),
      (c) -- [gluon] (a),
      (d) -- [gluon] (a),
      (e) -- [gluon] (a),
    };
  \end{feynman}
\end{tikzpicture}
\caption{On the left, the three-gluon vertex. On the right, the four-gluon vertex in Yang-Mills theory. Similar diagrams are found for W-bosons as well.}
\end{subfigure}
\\
\vspace{0.5cm}
\begin{subfigure}[b]{0.9\linewidth}
\centering
\includegraphics[width=0.7\linewidth]{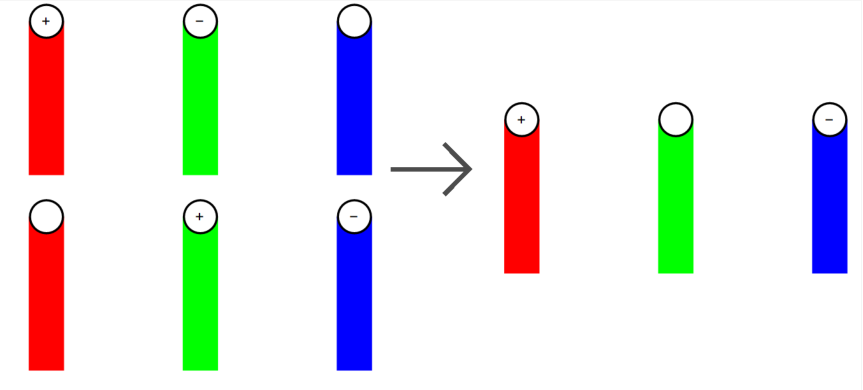}
\caption{The Bilson-Thompson braids for two non-Cartan gluons combine together to give a third gluon.}
\end{subfigure}
\caption{The gluon-interaction Feynman rules are shown on the top and a gluon interaction in the helon model is shown on the bottom.} 
\label{fig:3and4gluon}
\end{figure}

\begin{figure}[h!]
\centering

\begin{subfigure}[b]{\linewidth}
\centering
\begin{tikzpicture}[baseline={(current bounding box.center)}]
  \begin{feynman}
    \vertex (a);
    \vertex [above left=of a] (b) {\(W^{+}\)};
    \vertex [below left=of a] (c) {\(W^{-}\)};
    \vertex [right=of a] (d) {\(Z\)};

    \diagram* {
      (b) -- [boson] (a),
      (c) -- [boson] (a),
      (a) -- [boson] (d),
    };
  \end{feynman}
\end{tikzpicture}
\hspace{1.5cm}
\begin{tikzpicture}[baseline={(current bounding box.center)}]
  \begin{feynman}
    \vertex (a);
    \vertex [above left=of a] (b) {\(W^{+}\)};
    \vertex [below left=of a] (c) {\(W^{-}\)};
    \vertex [right=of a] (d) {\(\gamma\)};

    \diagram* {
      (b) -- [boson] (a),
      (c) -- [boson] (a),
      (a) -- [photon] (d),
    };
  \end{feynman}
\end{tikzpicture}

\subcaption{Feynman diagrams for \(W^{+}W^{-}Z\) (left) and \(W^{+}W^{-}\gamma\) (right) vertices.}
\end{subfigure}

\vspace{1em}

\begin{subfigure}[b]{\linewidth}
\centering
\includegraphics[width=0.45\linewidth]{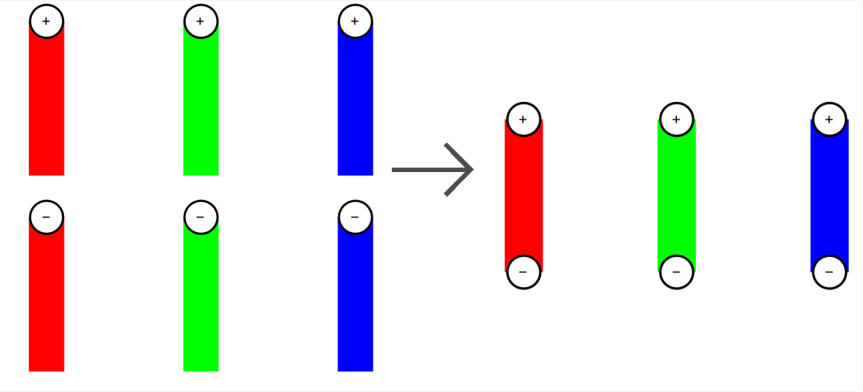}
\hspace{1cm}
\includegraphics[width=0.45\linewidth]{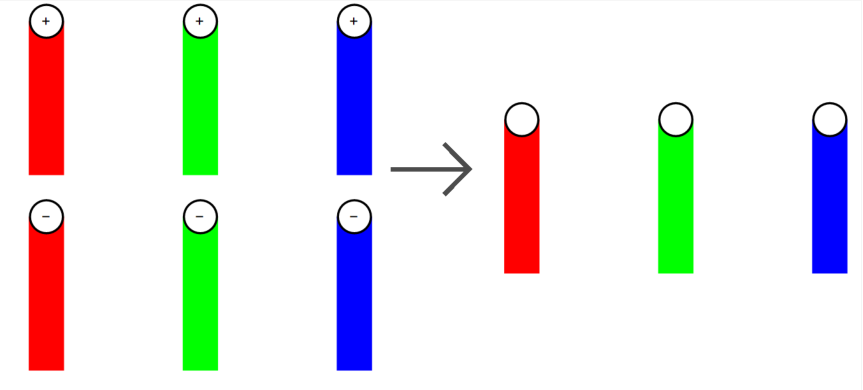}

\subcaption{Helon braid diagrams corresponding to the electroweak interactions.}
\end{subfigure}

\caption{The $W^+$ and $W^-$ bosons are shown to interact with the $Z$ boson and photon.}
\label{fig:electroweak}
\end{figure}

By setting $\partial_\mu = i k_\mu$ and keeping track of which field it operates on, the Yang-Mills action can be expanded to give
\begin{equation}
\mathcal{L}_A = \frac{1}{2}k_\mu A_\nu^A \left(k^\mu A^{\nu A} - k^\nu A^{\mu A}\right) -\frac{i}{2} k_{[\mu} A_{\nu]}^A gf^{ABC} A^{\mu B} A^{\nu C} - \frac{1}{4} g^2 f^{ABC} f^{ADE} A_\mu^B A_\nu^C A^{\mu D} A^{\nu E}.
\end{equation}
Varying this action with respect to three and four fields leads to the interaction rules for Feynman diagrams,
\begin{align}
\frac{\partial^3 \mathcal{L}_A}{\partial A_\mu^A(k_1) \partial A_\nu^B(k_2) \partial A_\rho^C(k_3) } =& -ig f^{ABC} \left[(k_1-k_2)^\rho \eta^{\mu\nu} + (k_2-k_3)^\mu \eta^{\nu\rho} + (k_3-k_1)^\nu \eta^{\rho\mu} \right], \\
\frac{\partial^4 \mathcal{L}_A}{\partial A_\mu^A(k_1) \partial A_\nu^B(k_2) \partial A_\rho^C(k_3) } =& -g^2\left[f^{ABE}f^{CDE}(\eta^{\mu\rho}\eta^{\nu\sigma} - \eta^{\mu\sigma}\eta^{\nu\rho}) \right. \nonumber \\
& \left. + f^{ACE}f^{BDE}(\eta^{\mu\nu}\eta^{\rho\sigma} - \eta^{\mu\sigma}\eta^{\rho\nu}) + f^{ADE}f^{BCE}(\eta^{\mu\nu}\eta^{\rho\sigma} - \eta^{\mu\gamma}\eta^{\sigma\nu}) \right] .
\end{align}
These two terms correspond to the Feynman diagrams shown in Figs.~\ref{fig:3and4gluon} and \ref{fig:electroweak}. To verify if these interactions are satisfied in the Bilson-Thompson model, we see if it is possible for two Yang-Mills bosons to be deformed into one of another color. This should suffice, as it is understood that the four-point QCD Feynman diagram can be factorized into two three-point interaction terms as utilized for color-kinematics duality \cite{Bern:2008qj}. 

To verify the interactions of gluons amongst Bilson-Thompson braids, one must impose the relation that combining a positive and negative twist lead to zero twist. In this sense, the Bilson-Thompson model can encode vacuum gluon interactions. However, we currently have not developed the Cartan generators of QCD rigorously enough to verify these interactions. Nevertheless, such interactions are typically less intuitive and are included primarily in sums over color, so it is plausible that the Bilson-Thompson model can accurately encode all gluon-gluon interactions once a more rigorous treatment of scattering amplitudes in the Bilson-Thompson model is provided. 

\begin{figure}
\centering
\includegraphics[width=0.6\linewidth]{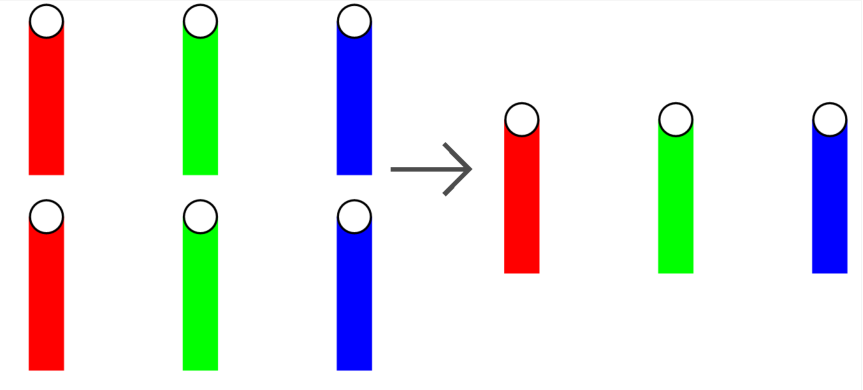}
\caption{Topologically, the combination of two photons could lead to a photon, which typically only occurs above the Schwinger limit and is not a classical phenomenon in the standard model. }
\label{fig:photon-fail}
\end{figure}

The terms above are ideally suited for QCD with massless gluons. Physics below the electroweak scale leads to massive $W^\pm$ and $Z$ bosons, while the photon and $Z$ bosons are mixtures of the $W^0$ and $B$ bosons. Therefore, additional care is warranted to verify the electroweak boson interactions. Rather than exhaustively reviewing further details of the electroweak sector, we focus directly on the Bilson-Thompson braids to point out some issues when dealing with electroweak gauge boson interactions:
\begin{enumerate}
\item The gluon interaction braid rules suggest a tension with the definition of the Z-boson. It is not clear how or why a Z-boson can contain strands that are simultaneously twisted and untwisted, especially since gluon interactions turn a twisted and untwisted strand into no twist. As such, the Z-boson is topologically equivalent to the photon. 
\item Since the photons are identity operators and the helon model suggests that combining braids are what encode interactions, it appears as if the photon can interact with itself, as depicted in Fig.~\ref{fig:photon-fail}. This issue may be avoided if a more comprehensive understanding of interactions of gauge bosons associated with Cartan vs non-Cartan generators is realized, as the interactions for the two Cartan generators of QCD was also less clear.
\item There is nothing in the helon model preventing W-bosons from interacting with right-chiral fermions. Since the $W^\pm$-bosons are adding a twist to each strand to the photon state and there is no braiding, the left and right-chiral fermions encoded by braids can freely act with all bosons equally in an achiral manner unless certain braid interactions are removed by hand.
\item There is no proposed Higgs boson in the helon model. Assuming a scalar boson has no braid states like the vector bosons, it would seemingly have to be identical to the photon, which makes it difficult to fit the Higgs field naturally in the helon model. 
\end{enumerate}

Overall, the helon model at a minimum requires additional constraints on which braids should not be allowed to be combined. The QCD sector of the helon mdoel appears to be the most successful, especially if the Cartan-generator gauge bosons can be described. The Z-boson seems topologically equivalent to the photon and the W-bosons seem to interact with both left- and right-chiral fermions. Besides complications with Cartan generator interactions (including the photon), it seems that the helon model can fully describe $SU(3)_c \times U(1)_{em}$ interactions, but not $SU(3)_c\times SU(2)_L\times U(1)_Y$. Since there is no candidate helon braid for the Higgs boson, there is no way to test those interactions, further suggesting that the helon model cannot currently describe the full standard model, yet does not rule out $SU(3)\times U(1)$. This point will be further clarified when studying boson-fermion-antifermion interactions next. 

\subsection{Interactions with Fermions}
\label{int-ferm}

\begin{figure}[h!]

\begin{subfigure}[t]{0.9\linewidth}

\centering
\begin{tikzpicture}[baseline=(current bounding box.center)]
  \begin{feynman}
    \vertex (a);
    \vertex [above left=of a] (b) {\(q\)};
    \vertex [below left=of a] (c) {\(\bar{q}\)};
    \vertex [right=of a] (d) {\(G_\mu^A\)};

    \diagram* {
     (b) -- [fermion] (a),
     (c) -- [anti fermion] (a),
     (a) -- [gluon] (d),
   };
  \end{feynman}
\end{tikzpicture}
\hspace{1cm}
\begin{tikzpicture}[baseline=(current bounding box.center)]
  \begin{feynman}
    \vertex (a);
    \vertex [above left=of a] (b) {\(q\)};
    \vertex [below left=of a] (c) {\(\bar{q}\)};
    \vertex [right=of a] (d) {\(W_\mu^i\)};

    \diagram* {
     (b) -- [fermion] (a),
     (c) -- [anti fermion] (a),
     (a) -- [boson] (d),
   };
  \end{feynman}
\end{tikzpicture}
\hspace{1cm}
\begin{tikzpicture}[baseline=(current bounding box.center)]
  \begin{feynman}
    \vertex (a);
    \vertex [above left=of a] (b) {\(q\)};
    \vertex [below left=of a] (c) {\(\bar{q}\)};
    \vertex [right=of a] (d) {\(B_\mu\)};

    \diagram* {
     (b) -- [fermion] (a),
     (c) -- [anti fermion] (a),
     (a) -- [boson] (d),
   };
  \end{feynman}
\end{tikzpicture}
\par\bigskip
\begin{tikzpicture}[baseline=(current bounding box.center)]
  \begin{feynman}
    \vertex (a);
    \vertex [above left=of a] (b) {\(l\)};
    \vertex [below left=of a] (c) {\(\bar{l}\)};
    \vertex [right=of a] (d) {\(W_\mu^i\)};

    \diagram* {
     (b) -- [fermion] (a),
     (c) -- [anti fermion] (a),
     (a) -- [boson] (d),
   };
  \end{feynman}
\end{tikzpicture}
\hspace{1cm}
\begin{tikzpicture}[baseline=(current bounding box.center)]
  \begin{feynman}
    \vertex (a);
    \vertex [above left=of a] (b) {\(l\)};
    \vertex [below left=of a] (c) {\(\bar{l}\)};
    \vertex [right=of a] (d) {\(B_\mu\)};

    \diagram* {
     (b) -- [fermion] (a),
     (c) -- [anti fermion] (a),
     (a) -- [boson] (d),
   };
  \end{feynman}
\end{tikzpicture}
\caption{Feynman diagrams for fermion-antifermion-gauge boson interactions. Quarks $q$, anti-quarks $\bar{q}$, leptons $l$, and anti-leptons $\bar{l}$ are shown above. Only the interactions that conserve color, weak isospin, and weak hypercharge are allowed.}
\end{subfigure}
\\
\vspace{0.5cm}
\begin{subfigure}[b]{\linewidth}
\centering

\includegraphics[width=0.45\linewidth]{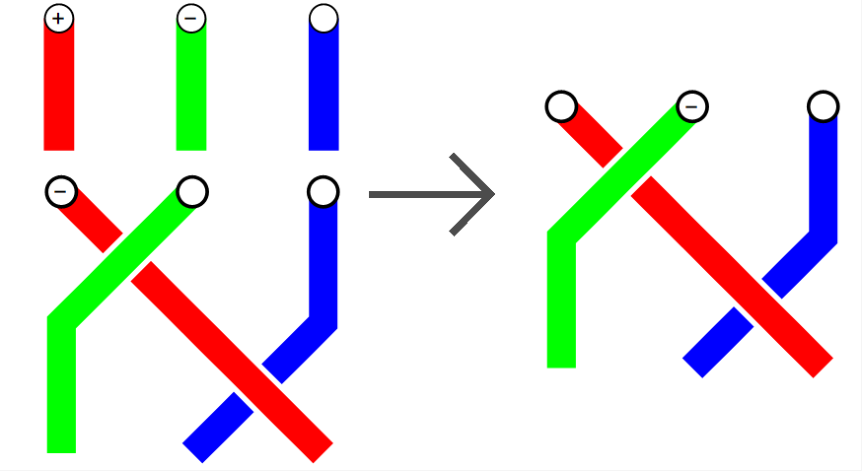}
\hspace{1cm}
\includegraphics[width=0.45\linewidth]{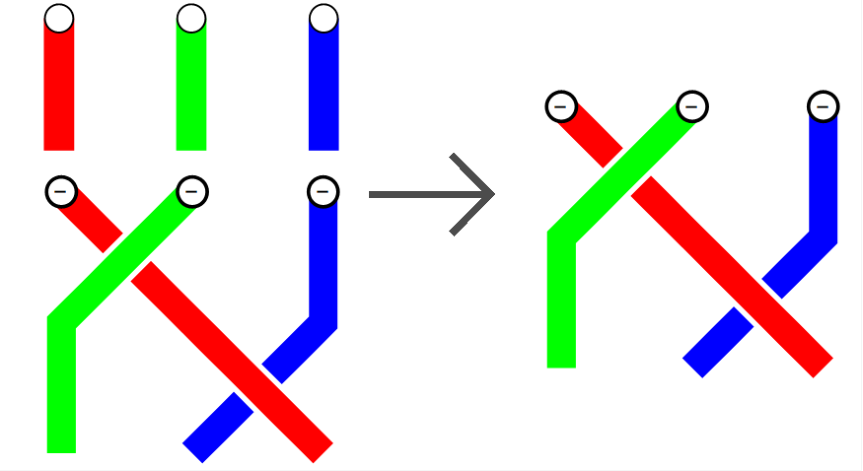}
\\
Down quark changes color from gluon \hspace{1cm} Electron with photon gives an electron
\\
\vspace{0.5cm}
\includegraphics[width=0.45\linewidth]{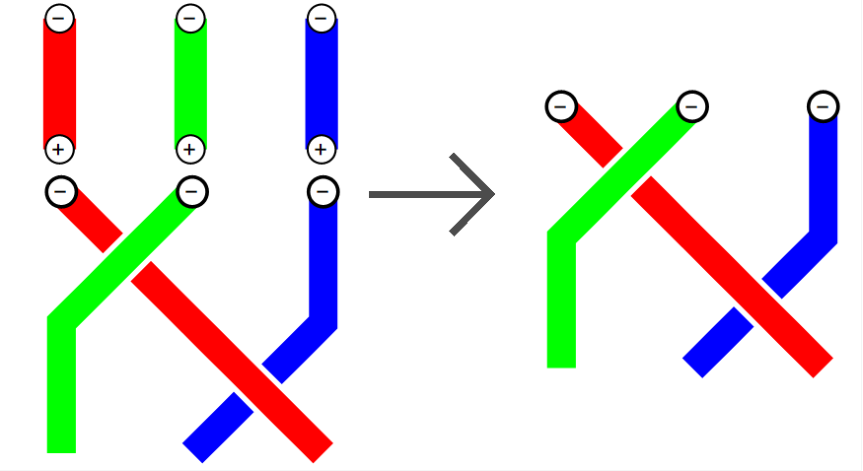}
\hspace{1cm}
\includegraphics[width=0.45\linewidth]{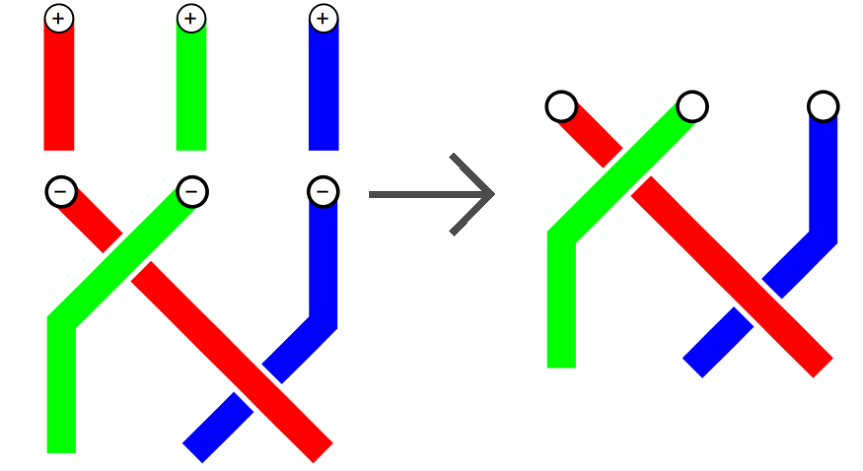}
\\
Electron with Z-boson gives an electron \hspace{0.75cm} Electron with W$^+$-boson gives a neutrino
\caption{Explicit examples of fermion/anti-fermion/gauge boson interactions are demonstrated by having a fermion braid incoming with the gauge boson braid to lead to an outgoing fermion braid. An example of a quark interacting with a $W$-boson is shown in Fig.~\ref{down-decay}.
}
\end{subfigure}
\caption{The fermion/anti-fermion/gauge boson interaction terms are depicted as Feynman diagrams and Bilson-Thompson braids above.}
\label{fig:fermion_gauge_interactions}
\end{figure}

The fermionic sector has interactions directly with the gauge bosons, which is encoded in the covariant derivatives. The Lagrangian for the different fields in momentum space are given by
\begin{align}
\mathcal{L}_q &= iq^\dagger \bar{\sigma}^\mu\left( ik_\mu  - i g_3 G_\mu  - i g_2 W_\mu - i Y_q B_\mu \right) q, \\ 
\mathcal{L}_l &= il^\dagger \bar{\sigma}^\mu\left( ik_\mu - i g_2 W_\mu - i Y_l B_\mu \right) l, \\ 
\mathcal{L}_ u &= iu^\dagger \bar{\sigma}^\mu\left( ik_\mu  - i g_3 G_\mu - i Y_u B_\mu \right) u, \\ 
\mathcal{L}_d &= ud^\dagger\bar{\sigma}^\mu\left( ik_\mu  - i g_3 G_\mu - i Y_d B_\mu \right) d, \\ 
\mathcal{L}_e &= ie^\dagger \bar{\sigma}^\mu\left( ik_\mu  - i Y_e B_\mu \right) e, \\ 
\mathcal{L}_\nu &= i\nu^\dagger \bar{\sigma}^\mu ik_\mu \nu.
\end{align}
The interaction rules for fermion-antifermion-gauge boson interactions are found by the following
\begin{align}
\frac{\partial^3 \mathcal{L}_q}{\partial q^\dagger_{\dot{\alpha} I i a} \partial q_{\beta J}^{j b} \partial G_\mu^A} = & g_3 (T^A)^i_{\,\,\, j} (\bar{\sigma}^\mu)^{\dot{\alpha}\alpha} \delta^a_b, \quad \frac{\partial^3 \mathcal{L}_q}{\partial q^\dagger_{\dot{\alpha} I i a} \partial q_{\beta J}^{j b} \partial W_\mu^i} =  g_2 (T^i)^a_{\,\,\, b} (\bar{\sigma}^\mu)^{\dot{\alpha}\alpha} \delta^i_j, \nonumber \\
\frac{\partial^3 \mathcal{L}_q}{\partial q^\dagger_{\dot{\alpha} I i a} \partial q_{\beta J}^{j b} \partial B_\mu} =&  Y_q g_1 (\bar{\sigma}^\mu)^{\dot{\alpha}\alpha} \delta^i_j \delta^a_b ,\\
\frac{\partial^3 \mathcal{L}_l}{\partial l^\dagger_{\dot{\alpha}Ia} \partial l_{\beta J}^{i}\partial W_\mu^k} = & g_2 (T^k)^a_{\,\,\, b} (\bar{\sigma}^\mu)^{\dot{\alpha}\alpha} \delta^i_j, \quad \frac{\partial^3 \mathcal{L}_l}{\partial l^\dagger_{\dot{\alpha}Ia} \partial l_{\beta J}^{i}\partial W_\mu^k} = Y_l g_1 (\bar{\sigma}^\mu)^{\dot{\alpha}\alpha} \delta^a_{b} \delta^i_j \\
\frac{\partial^3 \mathcal{L}_u}{\partial u^\dagger_{\dot{\alpha}Ii} \partial u_{\beta J}^{b} \partial G_\mu^A} =& g_3 (T^A)^i_{\,\,\, j} (\bar{\sigma}^\mu)^{\dot{\alpha}\alpha} \delta^a_b , \quad \frac{\partial^3 \mathcal{L}_l}{\partial u^\dagger_{\dot{\alpha}Ii} \partial u_{\beta J}^{j}\partial W_\mu^k} =  Y_u g_1 (\bar{\sigma}^\mu)^{\dot{\alpha}\alpha} \delta^a_{b} \delta^i_j  \\
\frac{\partial^3 \mathcal{L}_d}{\partial d^\dagger_{\dot{\alpha}Ii} \partial d_{\beta J}^b \partial G_\mu^A} =& g_3 (T^A)^i_{\,\,\, j} (\bar{\sigma}^\mu)^{\dot{\alpha}\alpha} \delta^a_b , \quad \frac{\partial^3 \mathcal{L}_l}{\partial d^\dagger_{\dot{\alpha}Ii} \partial d_{\beta J}^{b}\partial W_\mu^k} =  Y_d g_1 (\bar{\sigma}^\mu)^{\dot{\alpha}\alpha} \delta^a_{b} \delta^i_j \\
\frac{\partial^3 \mathcal{L}_e}{\partial e^\dagger_{\dot{\alpha}I} \partial e_{\beta J} \partial B_\mu} = &  Y_e g_1 (\bar{\sigma}^\mu)^{\dot{\alpha}\alpha} \delta^a_{b} \delta^i_j .
\end{align}
Each of these non-zero terms relates to a different interaction rule for Feynman diagrams, as shown in Fig.~\ref{fig:fermion_gauge_interactions}. To test if these hold in the Bilson-Thompson model, one may take the fermion and vector boson braids as incoming particles and combine them to see if they lead to a braid associated with a valid outgoing fermion. 

\begin{figure}[h!]
\centering
\begin{subfigure}{\linewidth}
\centering
\includegraphics[scale=0.5]{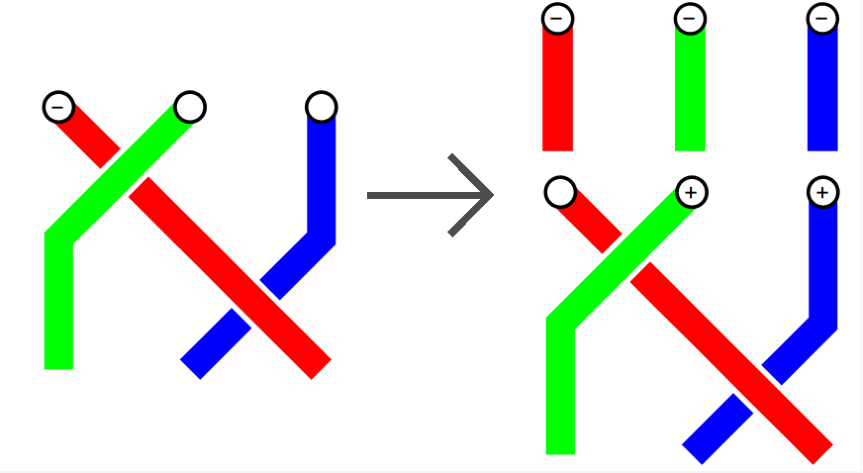}
\caption{Above, a decay process of $d_{L,r}\rightarrow W^- + u_{L,r}$ is depicted.
}
\vspace{0.4cm}
\end{subfigure}
\begin{subfigure}{\linewidth}
\centering
\includegraphics[scale=0.75]{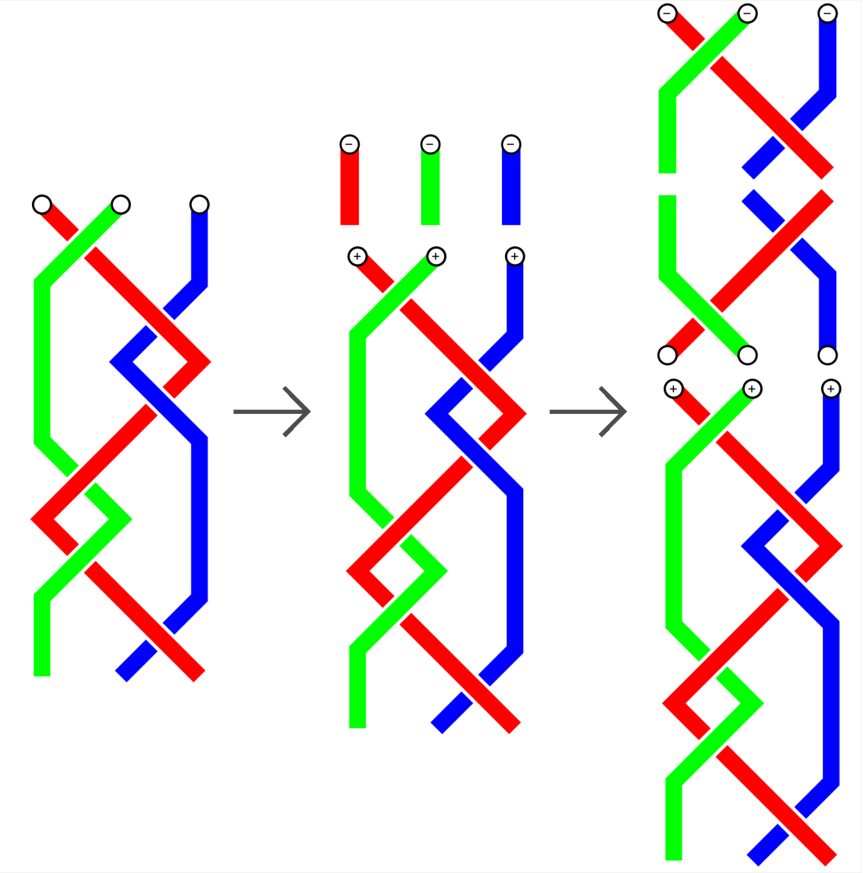}
\caption{
The neutron decay process is shown above with the following decay chain: $n \rightarrow W^- + p^+ \rightarrow e_{L} + \bar{\nu}_{R} + p^+$. 
}
\end{subfigure}
\caption{The anatomy of neutron decay with helon braids.}
\label{down-decay}
\end{figure}

A crucial modification to Bilson-Thompson's original particle assignment taken herein is the up quark. While Bilson-Thompson aligned a single braid operator $\sigma^{-1}_1 \sigma_2$ with the negatively charged states such as the electron and the anti-up quark, we focus on $\sigma_1^{-1} \sigma_2$ corresponding to left-chiral particles such as the left-chiral electron and up quark. This modification is necessary to adequately describe neutron decay. The neutron decay process contains an external incoming down quark decaying into an internal $W$-boson with an external outgoing up quark. As shown in Fig.~\ref{down-decay}, this interaction only works properly with our choice of the up quark. Additionally, neutron decay can be verified to work via topological means when considering quark composites from color-matched braid multiplication. However, it's worth reiterating that the helon model has no mechanism to stop a right-chiral weak decay process.

Besides the left-chiral nature of the weak force and the topological equivalence of the Z-boson with the photon, the helon model allows for all fermion-antifermion-gauge boson interaction terms. Additionally, the model does not encode the weak hypercharges, so these would have to be put in by hand. In this sense, the helon model can accommodate the $SU(3)_c \times U(1)_{em}$ gauge bosons and their interactions with all fermions found in the standard model. However, details of the electroweak sector, including the left-chiral nature of the weak force and the weak hypercharges are overlooked. This most likely relates to a misinterpretation of the braid model's gauge group. A similar pattern was found in Cohl Furey's exploration from $\mathbb{C}\otimes\mathbb{O}$ to $\mathbb{C}\otimes\mathbb{H}\otimes\mathbb{O}$ \cite{Furey:2010fm,Furey:2016ovx}, where an additional quaternionic structure was needed to be added to $\mathbb{C}\otimes \mathbb{O}$ to obtain $SU(3)\times SU(2)\times U(1)$ instead of $SU(3)\times U(1)$. To obtain the fully unbroken standard model with correct electroweak gauge boson interactions, Furey and Hughes were able to construct explicit fields and Lagrangians by using $\mathbb{C}\otimes\mathbb{H}\otimes \mathbb{O}$, which also accommodated a description of the Higgs field \cite{Furey:2022qhg}.

The interaction rules above allow for flavor changes in weak isospin from the $W^\pm$ bosons, but these interaction rules do not give rise to flavor changes across generations. However, classical (tree-level) Feynman diagrams can lead to cross-generational flavor changes through the weak force since the external legs are typically expressed in terms of mass eigenstates, while all interaction vertices are in terms of flavor eigenstates. Additionally, Feynman diagrams corresponding to quantum corrections have internal lines in mass eigenstates for particles in loops. 
The Yukawa sector primarily is for mass generation and describes the relationship between mass and flavor eigenstates through the CKM and PMNS matrices. Since the origin of mass and the Cabbibo mixing is not yet understood in the Bilson-Thompson model, there is no way to confirm if the Bilson-Thompson model agrees with the predictions from the Yukawa sector.

Finally, the helon model fails to capture the spin of the gauge bosons, while the braid diagrams do capture the fermionic spin states. We articulate that this may not be a significant problem with the model, but rather a clue towards how the model should be connected to the spacetime manifold. General relativity contains a global spacetime manifold, while Yang-Mills theory contains Minkowski spacetime with a local gauge group $G$. When studying gauge gravity, such as the work of Ivanov and Niederle \cite{Ivanov:1981wm}, it is customary to consider a global manifold with a local gauge group, such that $Spin(3,1) \subset G$. The gauge bosons such as the spin connection and the frame field are globally spin-1 fields, which gets their vector index directly from the global manifold. In this formulation of gauge gravity, the entire matter sector is treated as a scalar with respect to the global manifold. As such, fermions obtain their spin from the representation theory of the local gauge group, which differs from how the gauge bosons obtain their spin. It appears as if the helon model is connected to the representation theory of the local gauge group $Spin(3,1)\times SU(3)_c \times U(1)_{em}$, but not the global spacetime manifold. This could describe why the helon model provides the chirality of the fermions, but not the chirality of the gauge bosons, as the gauge bosons obtain their spin from the global manifold, not the local spacetime group for gauge gravity like the fermions do.

\section{The Helon Model from the $SU(3)_c\times U(1)_{em}$ Weight Lattice}
\label{weight-lattice}


The helon model completely describes the $SU(3)_c\times U(1)_{em}$ gauge theory for a single generation of standard model fermions. Next, we seek to understand the origin of these braid and twist assignments from representation theory. 
Note that the on-shell states of  spinors are  determined by the braiding information. Therefore, the twists can then uniquely determine the charge states of  different elementary particles. As we will show, the twist values of the helon braids can be found from an affine linear transformation acting on the weight lattice coordinates. 

\begin{figure}[h!]
\centering
\includegraphics[scale=0.9]{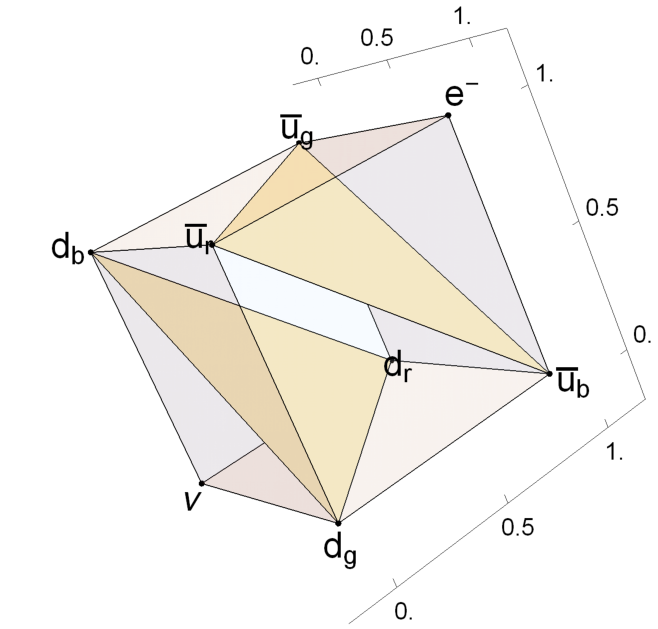}
\caption{The twist values of the three strands for the non-negatively charged fermions in the first generation are show above to give a cube.}
\label{fermcube}
\end{figure}

The fact that the ribbon twists determine the electric charge is further evidence that the helon model encodes $SU(3)_c\times U(1)_{em}$ rather than $SU(3)_c\times SU(2)\times U(1)_Y$. In the helon model, each ribbon can be untwisted, twisted clockwise, or twisted counter-clockwise, which can be encoded by the integers 0, 1, and -1, respectively. When taking these values as 3D coordinates, the fermionic states can be viewed as two cubes sharing a corner at the origin. The fermions with positive electric charge contain positive coordinates and are vertices of one cube, while fermions with negative electric charge are encoded by vertices on the other cube. The neutrino is found at the origin. The two cubes are given by the following coordinates,
\begin{align}
(0,0,0), \nonumber \\ 
(\pm 1, 0, 0), (0,\pm 1, 0), (0,0,\pm 1), \label{twistCoords}\\ 
(\pm 1, \pm 1, 0), (\pm 1, 0, \pm 1), (0, \pm 1, \pm 1), \nonumber\\ 
(\pm 1, \pm 1, \pm 1), \nonumber
\end{align}
where taking slices across the body diagonal of the cube isolates four distinct geometric objects as two points and two equilateral triangles. These correspond to the $SU(3)_c$-singlet leptons and the triplet quarks, respectively. Fig.~\ref{fermcube} depicts a cube and provides the particle assignments. 

While the relationship of the $A_2$ root system and $SU(3)$ is well established, $U(1)$ does not lead to a root system. The dimension of a root lattice is determined by the maximal number of commuting generators, often called Cartan generators. Each root vector corresponds to a non-Cartan generator. All of the representations of a Lie group are encoded in the weight lattice, which is closely related to the root lattice. Representations correspond to a collection of points often called weight polytopes. Cartan's classification of simple Lie groups in terms of root systems does not include $U(1)$ because $U(1)$ is not a simple Lie group. Nevertheless, $U(1)$ has a single Cartan generator, since it commutes with itself. This is identical to $A_1$ associated with $SU(2)$, whose Cartan generator is also a generator of $U(1)$. Additionally, when branching from $SU(2)$ to $U(1)$, the representations of $SU(2)$ lead to $U(1)$ singlets with different charges found from the coordinates of the $A_1$ weight lattice. 

For this reason, even though $U(1)$ has no formal root system, representations with $U(1)$ charges can be placed on an $A_1$ weight lattice. It is common to find $U(1)$ charges as subgroups of a larger group $G$ by projecting from the root lattice or weight lattice of $G$. Cohl Furey also found a similar structure with her study of $\mathbb{C}\otimes \mathbb{O}$ to model $SU(3)_c\times U(1)_{em}$ interactions \cite{Furey:2010fm,Furey:2016ovx}. The existence of such a fermionic cube suggests that a rotation and scale transformation can relate the ribbon twists of the helon model to the weight lattice coordinates as two different bases of the same space. 

First, the $A_2$ weight lattice coordinates of relevant $SU(3)$ representations are reviewed. Then, we introduce the $U(1)$ charges as an extra dimension associated with an independent $A_1$ weight lattice. Next, the $A_2\oplus A_1$ weight lattice is rescaled such that two cubes are found for the fermionic states with $SU(3)_c\times U(1)_{em}$ gauge symmetry. Finally, a scale and rotation transformation are found to rotate the weight polytopes in the cubes to the twist coordinates found in the helon model.

\begin{figure}[h!]
\centering
\includegraphics[scale=0.7]{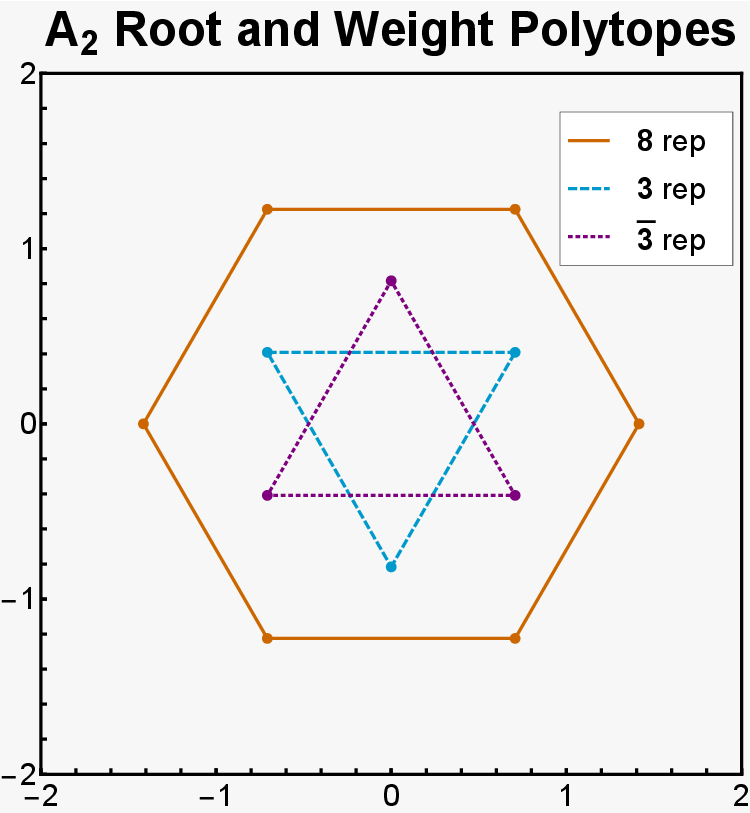}
\caption{The root vectors of $A_2$ form a hexagon in the root lattice, while the fundamental and anti-fundamental representations correspond to weight vectors forming equilateral triangles in the weight lattice.}
\label{A2poly}
\end{figure}

The $A_2$ root polytope is given by a hexagon, which corresponds to the six non-Cartan generators of $SU(3)$ in the adjoint representation. These six non-Cartan generators are combined with the two Cartan generators of $A_2$ to give the eight generators of $SU(3)$ in the ${\bf8}$ representation. The $A_2$ weight lattice is more dense, such that the ${\bf3}$ and $\overline{{\bf3}}$ representations are contained as two dual equlateral triangles. Since the tensor product ${\bf3}\otimes \overline{{\bf3}} = {\bf8} \oplus {\bf1}$, the $A_2$ root polytope can also be found in the weight lattice by adding together two weight vectors associated with ${\bf3}$ and $\overline{{\bf3}}$. The singlet state ${\bf 1}$ corresponds to the origin in the weight lattice. It is conventional to normalize the root vectors to have a length of $\sqrt{2}$, which leads to the weight vectors with a length of $\sqrt{\frac{2}{3}}$. The $A_2$ polytopes are shown in Fig.~\ref{A2poly}. 

The standard model fermions can be expressed as representations of $SU(3)_c\times U(1)_{em}$, such as
\begin{equation}
e: {\bf1}_{-1}, \qquad \bar{u}: \overline{{\bf3}}_{-2/3}, \qquad d: {\bf3}_{-1/3}, \qquad \nu: {\bf1}_0.
\end{equation}
By taking the 2D coordinates of the representations in the weight lattice and appending the electric charges, the following coordinates are found,
\begin{align}
e: & \,\, (0,0,-1), \label{ecoords} \\ 
\bar{u}_i: &\,\, \left\{\left(0,\sqrt{\frac{2}{3}},-\frac{2}{3}\right), \left(\frac{1}{\sqrt{2}},-\frac{1}{\sqrt{6}},-\frac{2}{3}\right), \left(-\frac{1}{\sqrt{2}},-\frac{1}{\sqrt{6}},-\frac{2}{3}\right)\right\}, \\
d_i: &\,\, \left\{\left(0,-\sqrt{\frac{2}{3}},-\frac{1}{3}\right), \left(-\frac{1}{\sqrt{2}},\frac{1}{\sqrt{6}},-\frac{1}{3}\right), \left(\frac{1}{\sqrt{2}},\frac{1}{\sqrt{6}},-\frac{1}{3}\right)\right\}, \\
\nu: &\,\, (0,0,0). \label{nucoords}
\end{align}
The coordinates above provide a deformed cube that is squished along the body diagonal, which corresponds to the $A_1$ weight lattice due to the electric charge. Since the $A_2$ and $A_1$ are independent, we can rescale the $A_2$ weight lattice by a factor of $1/\sqrt{3}$ to form a cube. 

Since the twist coordinates and the rescaled weight lattice coordinates both give a cube with a vertex at the origin, a final rotation and scale transformation can find the twist coordinates from the weight lattice coordinates.
To obtain the twist values from the weight lattice coordinates with $A_2$ roots normalized to $\sqrt{2}$ and the $A_1$ lattice scaled to the electric charge values, the transformation matrix is given by
\begin{equation}
M = \left(\begin{array}{ccc} -\frac{1}{\sqrt{2}}&-\frac{1}{\sqrt{6}}&\frac{1}{\sqrt{3}} \\ \frac{1}{\sqrt{2}} &-\frac{1}{\sqrt{6}}&\frac{1}{\sqrt{3}} \\ 0&\sqrt{\frac{2}{3}}&\frac{1}{\sqrt{3}} \end{array} \right)  \left(\begin{array}{ccc} \sqrt{3}&0&0 \\ 0&\sqrt{3}&0 \\ 0&0&\sqrt{3} \end{array} \right)  \left(\begin{array}{ccc} \frac{1}{\sqrt{3}} & 0 & 0 \\ 0 & \frac{1}{\sqrt{3}} & 0 \\ 0 & 0 & 1 \end{array}\right)  = \left(\begin{array}{ccc} -\frac{1}{\sqrt{2}}&-\frac{1}{\sqrt{6}}&1 \\ \frac{1}{\sqrt{2}}&-\frac{1}{\sqrt{6}}&1 \\ 0&\sqrt{\frac{2}{3}}&1 \end{array} \right).
\end{equation}
Applying the matrix $M$ above to Eqs.~\eqref{ecoords}-\eqref{nucoords} leads to the non-positive coordinates shown in Eq.~\eqref{twistCoords}. This demonstrates that the twist values of the helon model only contain sufficient information to encode the charges (and interactions) of $SU(3)_c \times U(1)_{em}$, not $SU(3)_c\times SU(2)_L \times U(1)_Y$ as originally claimed. 

Next, the goal is to understand if the braid operators $\sigma_1$, $\sigma_2$, $\sigma_1^{-1}$, and $\sigma_2^{-1}$ can be identified in terms of weight lattice coordinates associated with representations of the little group Lorentz group, as these correspond to the on-shell states for the fermions. While the Lorentz group is associated with the $D_2$ root system, the massless little group of $Spin(3,1)$ is $Spin(2)$, while the massive little group is $Spin(3)$. Since the helon model contains no Higgs field, we consider the masses to be put in, implying $Spin(3) \cong SU(2)$ associated with $A_1$ should be chosen. Even if $Spin(2) \cong U(1)$ was chosen, this would be uplifted to the $A_1$ root system, as done with $U(1)_{em}$. This implies that the data in the helon braids should be recoverable from 4D coordinates that add an additional dimension to the 3D coordinates discussed above for the twist. 

A Dirac bispinor contains two irreducible representations associated with left and right chiral spinors, which is often depicted as $\left(\frac{1}{2},0\right) \oplus \left(0,\frac{1}{2}\right)$. In terms of weight lattice coordinates, $(1,0)$, $(1,0)$, $(0,1)$, and $(0,-1)$ would be obtained for this bispinor, which also can be identified as $D_2 \cong A_1 \oplus A_1$. Since these are complex representations of $Spin(3,1) \cong SL(2,\mathbb{C})$, the 8 off-shell degrees of freedom include complex roots. The on-shell states of a Dirac bispinor therefore are specified by coordinates of a 2D weight lattice, not a 1D weight lattice. This can be understood because a massless Dirac spinor is an irreducible representation of the conformal group $Spin(4,2)\cong SU(2,2)$ associated with $D_3 \cong A_3$. This makes it appear as if 5D coordinates would be required to encode the helon model. 

Curiously, the $\left(\frac{1}{2},0\right)$ states always put $\sigma_1^{-1}$ or $\sigma_2^{-1}$ first, while $\left(0,\frac{1}{2}\right)$ always puts $\sigma_1^{-1}$ or $\sigma_2^{-1}$ last. $\sigma_1^{-1}$ represents a particle, while $\sigma_2^{-1}$ represents an antiparticle. The position $\sigma_1^{\pm 1}$ signifies whether the particle is left- or right-chiral. This self-consistency for all braids provides further support that the braid model is a faithful characterization of the chiral fermion states.

\begin{figure}
\centering
    \includegraphics[scale=0.7]{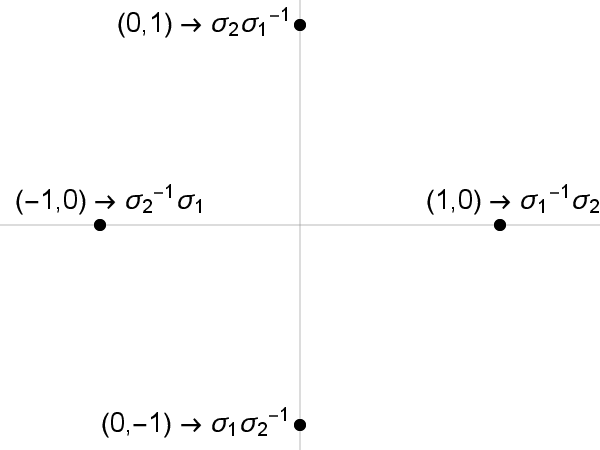}
    \caption{The coordinates for the $D_2$ weight lattice and the corresponding braid operators are shown above.}
    \label{braid-diamond}
\end{figure}

Combining the four on-shell spin states represented by 2D coordinates with the 3D coordinates for electric and color charge leads to 5D coordinates for $D_2 \oplus A_2 \oplus A_1$ associated with $SL(2,\mathbb{C})\times SU(3) \times U(1)$. While the braid operators relate to the first two coordinates, the twist operators $T_i$ and $T_i^{-1}$ for $i=1,2,3$ correspond to the last three coordinates in 5D. The states of all positively twisted braids include the operators $T_{i}$, $T_iT_j$, and $T_1 T_2T_3$. These 5D weight lattice coordinates correspond to the following twisted braid states,
\begin{equation}
\begin{array}{ccc}
  (0,1,1,0,0)\rightarrow \sigma_2\sigma_1^{-1} T_1 , & (0,1,0,1,0)\rightarrow \sigma_2\sigma_1^{-1} T_2 , & (0,1,0,0,1)\rightarrow \sigma_2\sigma_1^{-1} T_3 , \\
  (-1,0,1,1,0)\rightarrow \sigma_2^{-1}\sigma_1 T_1 T_2 , & (-1,0,0,1,1)\rightarrow \sigma_2^{-1}\sigma_1 T_2 T_3^ , &  (-1,0,1,0,1)\rightarrow \sigma_2^{-1}\sigma_1 T_3 T_1, \\
    & (0,1,1,1,1)\rightarrow \sigma_2\sigma_1^{-1} T_1  T_2T_3.& 
  \end{array}
\end{equation}

Since the 5D coordinates are derived from the weight lattice, addition is guaranteed to respect sensible particle interactions for the trivial, fundamental, and adjoint representations. When two root vectors are added, they in general do not add to give another root vector \cite{humphreys1972}. However, when two dissimilar root vectors are added, they always lead to another root vector or the origin. 
This ensures that the addition of two gluons can lead to another gluon as shown in Fig.~\ref{fig:3and4gluon}. However, the helon model does not allow for two of the same gluons to interact, as this leads to strands with more than one twist. Similarly, it would not make sense to add two of the same root vectors and expect to get another root vector.

Similarly, in the weight lattice, the quark states can be flipped from red to green to blue by adding various gluon states identified in the weight lattice. Since the standard model does not include any higher-dimensional representations beyond the adjoint representation, there are no particle states found from adding two quarks of the same color. However, a quark and an anti-quark either lead to a color/anti-color state such as a gluon or they lead to a color singlet such as a photon. As such, a partial binary addition operator can model particle interactions in the weight lattice. 

From the perspective of $SU(3)\times SU(2)\times U(1)$, only left-chiral spinors are chosen, since the Higgs mechanism is used to give the fermions mass. Therefore, if the standard model gauge group was accounted for by the helon model, then it appears that only a single coordinate would be required to identify particle vs antiparticle associated with the $\left(\frac{1}{2},0\right)$ representation of the Lorentz group. This agrees with the notion that the little group of $Spin(3,1)$ would correspond to a 1D weight lattice, which should uniquely encode the on-shell states of an irreducible representation of the Lorentz group. Additional care would be needed to recognize that the right-chiral fermions are treated as left-chiral charge-conjugated states. However, the weight lattice of $SU(3)\times SU(2)\times U(1)$ would lead to $A_2\oplus A_1 \oplus A_1$, so 5D coordinates still seem necessary in this picture. This begs the question of whether it is possible to use 4D coordinates to recover all of the information in the helon model. 

Lambek's 4D model was also inspired by Harari and Shupe, which used addition of quaternions to encode interactions. Lambek's work also has similar claimed assignments for the Z- and W-bosons as Bilson-Thompson, but Lambek has the correct states for the up quark. However, Lambek's 4D model did not account for chirality. Instead, the fourth ``scalar'' dimension added refers to particle vs antiparticle. As such, Bilson-Thompson's model has double the braid states as Lambek's 4-vector states. Lambek's 4D model can be recovered from the corrected helon model by removing chirality.  Lambek's 4D model can also be found via projection from the 5D coordinates via $(x+y,r,g,b)$ for 5D coordinates $(x,y,r,g,b)$. This projection loses the information about chirality and therefore cannot distinguish all of the on-shell fermionic states. 

The helon model braids can, however, be uniquely determined from different 4D weight lattice coordinates $(y-x,r,g,b)$ that identifying chirality instead of particle/antiparticle. This is possible because the particle/antiparticle state in a single chiral spinor also flips the charges. Since the standard model has no mirror fermions, the electric charge can be used to uniquely determine if a fermionic state is a particle or an antiparticle. This allows for the chirality to be added as the 4th dimension to the 3D twist coordinates. One must resort to particle physics knowledge to reconstruct the braid operators from these 4D coordinates. It is not possible to recover the helon model from Lambek's 4D model because there is no way to determine if a state should be left-chiral or right-chiral from Lambek's 4D coordinates. The newly proposed 5D or 4D coordinates allow for the helon model to be recovered. Table \ref{5Dcoords} depicts the helon braids, Lambek's 4D model, the 5D weight lattice coordinates stemming from $SL(2,\mathbb{C})\times SU(3)_c \times U(1)_{em}$, and the 4D weight lattice coordinates with chirality. 

\begin{table}
\centering
\begin{tabular}{c||c|c|c|c}
Particle & Operator & 4D Lambek & 5D Weight Lattice  & 4D Weight Lattice \\ 
\hline \hline
$e_L$ & $\sigma_1^{-1} \sigma_2 T_{123}^{-1}$ & (1,-1,-1,-1) & (1,0,-1,-1,-1) & (-1,-1,-1,-1) \\ \hline  
$e_R$ & $\sigma_2 \sigma_1^{-1} T_{123}^{-1}$ & (1,-1,-1,-1) & (0,1,-1,-1,-1) & (1,-1,-1,-1) \\  \hline
$\bar{u}_{b,L}$ & $\sigma_1\sigma_2^{-1} T_{12}^{-1}$ & (-1,-1,-1,0) & (0,-1,-1,-1,0) & (-1,-1,-1,0) \\ \hline
$\bar{u}_{b,R}$ & $\sigma_2^{-1}\sigma_1 T_{12}^{-1}$ & (-1,-1,-1,0) & (-1,0,-1,-1,0) & (1,-1,-1,0) \\ \hline
$d_{r,L}$ & $\sigma_1^{-1}\sigma_2 T_{1}^{-1}$ & (1,-1,0,0) & (1,0,-1,0,0) & (-1,-1,0,0) \\ \hline
$d_{r,R}$ & $\sigma_2\sigma_1^{-1} T_{1}^{-1}$ & (1,-1,0,0) & (0,1,-1,0,0) & (1,-1,0,0) \\ \hline
$\nu_L$ & $\sigma_1^{-1}\sigma_2$ & (1,0,0,0) & (1,0,0,0,0) & (-1,0,0,0) \\ \hline
($\nu_R$) & $\sigma_2\sigma_1^{-1}$ & (1,0,0,0) & (0,1,0,0,0) & (1,0,0,0) \\
\hline\hline
\end{tabular}
\caption{The fermionic spectrum of the standard model is depicted in terms of their braid diagrams associated with $SU(3)_c \times U(1)_{em}$, their 4D Lambek coordinates, their 5D weight lattice coordinates associated with $D_2\oplus A_2 \oplus A_1$, and a 4D weight lattice that removes the distinction of particle/antiparticles. The non-interacting right-chiral neutrino can be included or not included in the model. For the twist operators, $T_{i\dots k}^{\pm 1} \equiv T_i^{\pm 1} \dots T_k^{\pm 1}$ as a shorthand.}
\label{5Dcoords}
\end{table}

\section{Conclusions and Discussion }
\label{conc}

In conclusion, this work started with a discussion of various preon models from  Harari and Shupe, to Lambek, to Bilson-Thompson. When preon models were first introduced in particle physics, they generated some excitement in the community due to the fact that their combinatorics could reproduce aspects of standard model phenomenology. However, it was never clear where these interaction rules came from, or how these different models mapped onto each other. The most recent among preon models are the 4-vector model and the helon model (which incorporate the data of earlier models). We have shown how particle states in the helon model map to the weight lattice of $SU(3)_c\times U(1)_{em}$. Braids correspond to on-shell states of spinors of the Lorentz group, whereas twists of helons denote charges. Braid topology as a representation of particle chirality was shown to be understood in terms of the mapping of the  braid group $\mathcal{B}_3$ to $SL (2, \mathbb{Z})$, which in turn, sits within $SL (2, \mathbb{C})$, the double cover of the restricted Lorentz group. Additionally, we have put forth an operational realization of C, P and T transformations in terms of braid topology and twists. Remarkably, in terms of these transformations, we found that the braid diagrams of the helon model are precisely the only ones that happen to be CPT invariant. 

In this work, we have established the complete mapping between the helon model and Lambek's 4-vector model. This includes the realization of composite particles in the helon model, which was unclear until now. With the color-matched braid multiplication proposed here, composite particles such as neutrons and protons can now be consistently expressed via helon model diagrams. In the 4-vector model these composite particles were easily expressible, but lacked chirality data. The revised helon model proposed here, now accounts for this.  

Moreover, we have worked out a new 5-vector representation fully derived from the Lie algebra weight lattice. As we have demonstrated in this work, this contains both the correct interactions found in Lambek and the inclusion of chirality found in Bilson-Thompson. Additionally, a new 4-vector model that differs from Lambek was also found to contain two chiral states of fermions. 

Now let us discuss a few outstanding issues and their possible resolutions. One is that the helon model does not precisely capture the left-handedness of the weak interactions. That is, compared to standard field theory formalisms, there is no projection operator that projects out right-handed components during a weak interaction process. Given that the full set of standard model interactions can not be found in the helon model, this motivates the question whether the weight polytope coordinates for $SU(3)_c \times SU(2)_L \times U(1)_Y$ should be used to find twist coordinates for a potential generalized helon model. Such a generalized helon model would stem from an $A_2 \oplus A_1 \oplus A_1$ lattice, which suggests additional strands to helon braids beyond three. For generalized helon models of arbitrary strands to work for on-shell fermions states, one would also require that they satisfy CPT invariance.   

Another issue concerns higher generation fermions. Note that one of the remarkable features of the helon model is that arbitrary combinations of braids are not CPT invariant (under the version of CPT transformations that have been articulated in terms of braid operations). Only braids corresponding to known quarks, leptons and bosons are CPT invariant. In other words, this model does not admit any spurious particle states. That being said, it is still not clear how to represent higher generation standard model fermions. The simplest possibility of representing these fermions with additional braid crossings turns out not to be consistent, since braid diagrams with even number of crossings greater than two already consistently express composites of particles (such as neutrons, protons or collections of quarks and leptons). Here again it seems an appropriate generalization involving braid groups beyond $\mathcal{B}_3$, that is with additional strands, may be called for. As mentioned earlier, such a generalization also has the potential to capture the left-handedness of the weak interactions. 

More generally, as alluded to by both, Dirac and Bilson-Thompson, topology serves as a useful description to understand the underlying rules governing fundamental interactions of particles, that is, the compositionality inherent in the fundamental forces of the standard model. Hence, rather than being structural extensions of fundamental particles in physical space, what preon models allude to are  combinatorial properties governing fundamental interactions. As we have shown here, these combinatorial properties are in fact, directly related to the underlying gauge group and their associated symmetries. However, while gauge group formalisms usually adopt a field-centric view, combinatorial preon models provide the analogous particle-centric perspective. It was in this sense, that Lambek referred to this approach as the "grammar" of particle interactions. Gauge field theories are typically  formulated in terms of the geometry  of principal $G$-bundles, defined by the group action of the corresponding gauge group $G$. There, one often regards the configuration space as the physical external space, whereas the fiber space, consists of internal degrees of freedom. Much like the spin of a fermion, where the internal degrees of freedom correspond to an isospin space. The crucial point is that all of these internal spaces have topological or geometric properties, which lead to observable consequences, even though the topological structure itself is not what one  "directly" observes. It is the compositionality of these topological structures corresponding to internal degrees of freedom (defined via underlying gauge symmetries) that preon models effectively capture.

\section*{Acknowledgements}

We thank  Cohl Furey and Mia Hughes for multiple discussions related to their work.  \\ It gives the third author pleasure to thank the Institute for Sustainability of Knotted Chiral Meta Matter ($SKCM^{2}$) at Hiroshima University, Hiroshima, Japan for hospitality and resources during the preparation of this research.

\appendix

\section{Standard Model Conventions }
\label{std-model}

In this appendix, we review some well-established facts about  $SU(3)_c \times SU(2)_L \times U(1)_Y$ gauge theory, in order to establish the conventions used in the main text. This overview can be found in standard texts such as \cite{weinberg1995quantum}. 

$SU(3)_c \times SU(2)_L \times U(1)_Y$ gauge theory contains a spectrum of particles described by quantum fields under the representations of $SL(2,\mathbb{C})\times SU(3) \times SU(2) \times U(1)$ once including the spin degrees of freedom. By using the chiral representations of the Lorentz group, the $(\textbf{p},\textbf{q})$ representation of $SL(2,\mathbb{C})$ has left-chiral spinors as $(\textbf{2},\textbf{1})$, vector fields as $(\textbf{2},\textbf{2})$, and scalars as $(\textbf{1},\textbf{1})$. Note that technically, a Weyl spinor in $D=3+1$ is a complex spinor, which implies that the complex conjugate state is also included. In other words, a left-chiral particle also contains degrees of freedom for the CPT-conjugate right-chiral anti-particle $(\textbf{1},\textbf{2})$, which would always have opposite charges.

\begin{table}
\centering
\begin{tabular}{c||c|c|c|c}
Group & $SL(2,\mathbb{C})$ & $SU(3)_c$ & $SU(2)_L$ & $Spin(3)_F$ \\
\hline \hline
Fundamental & $\alpha$ & $i$ & $a$ & $I$ \\ 
\hline
Adjoint & $[ab]$ & $A$ & $i$ & $[IJ]$
\end{tabular}
\label{indices}
\caption{Indices for various reprentations of Lie groups for spacetime and the standard model. Note that $\alpha = 1,2$, $a=1,2$, $i=1,2,3$, $I=1,2,3$, and $A=1,\dots 8$. }
\end{table}

The standard model contains three generations of fermions, including the left-chiral quarks $q^{i a}$, the left-chiral leptons $l^a$, and the left-chiral charge conjugate of the up quark $u^i$, down quark $d^i$, and electron $e$, generalized to three families. While the standard model was defined to have massless neutrinos, it is now understood that at least two generations of neutrinos are massive. The status of right-chiral neutrinos is unknown. Since the braids for right-chiral neutrinos are natural, we consider the possibility of right-chiral neutrinos and will remain agnostic about the details of mass assignment to see if the helon model provides any constraints or clues. The bosons include the 12 vector gauge bosons and the electroweak Higgs field, which lead to the gluons $g^A$, the W-bosons $W^\pm$, the $Z$-boson $Z$, and the photon $A$ as gauge vector bosons and the Higgs field $H$ as a scalar below the electroweak scale.

The interactions of the standard model can be described by Feynman diagrams, whose interaction rules can be read off from the action $S$ or Lagrangian (density) $\mathcal{L}$. The Lagrangian for the standard model can be broken into the following pieces,
\begin{equation}
\mathcal{L}_{SM} = \mathcal{L}_{gauge} + \mathcal{L}_{fermion} + \mathcal{L}_{Higgs} + \mathcal{H}_{Yukawa}.
\end{equation}
The vector boson Lagrangian contains the $SU(3)\times SU(2) \times U(1)$ gauge bosons,
\begin{equation}
\mathcal{L}_{gauge} = -\frac{1}{4}tr\left(G_{\mu\nu}G^{\mu\nu} + W_{\mu\nu}W^{\mu\nu} + B_{\mu\nu}B^{\mu\nu}  \right),
\end{equation}
where the gluon field strength $G_{\mu\nu}$, the weak field strength $W_{\mu\nu}$, and the weak hypercharge field strength $B_{\mu\nu}$ are matrix fields with respect to the $SU(3)$, $SU(2)$, and $U(1)$ gauge fields, respectively. The vector potentials are given by $G_\mu = G_\mu^A T_3^A$, $W_\mu = W_\mu^i T_2^i$, and $B_\mu$,
\begin{align}
G_{\mu\nu} &= G_{\mu\nu}^A T^A = \partial_\mu G_\nu - \partial_\nu G_\mu + [G_\mu, G_\nu], \\ 
W_{\mu\nu} &= W_{\mu\nu}^i T^i = \partial_\mu W_\nu - \partial_\nu W_\mu + [W_\mu, W_\nu], \\ 
B_{\mu\nu} &= \partial_\mu B_\nu - \partial_\nu B_\mu.
\end{align}
Upon spontaneous symmetry breaking of $SU(2)_L \times U(1)_Y \rightarrow U(1)_{em}$, the $W^0$- and B-bosons get mixed to give the $Z$-boson and photon. The vacuum expectation value (vev) of the electroweak Higgs relates to the Higgs mass. Weinberg's mixing angle rotates the $W^0$ and $B$ bosons to the $Z$ and $\gamma$ bosons, which also relates to the difference of the mass of the Z and $W^\pm$ bosons. The gluons, W-bosons, and B-boson correspond to the following representations of $SU(3)\times SU(2)\times U(1)$: $(\textbf{8},\textbf{1})_0 \oplus (\textbf{1},\textbf{3})_0 \oplus (\textbf{1},\textbf{1})_0$. The Higgs field is in the $(\textbf{1},\textbf{2})_{1/2}$ representation.

The fermionic Lagrangian can be expressed either all as left-chiral fields or a mix of left-chiral weak isospin doublets and right-chiral singlets. Complex conjugation allows for the two to be related. Since the $\textbf{16}$ spinor of $Spin(10)$ GUT unifies the first generation of fermions with the right-chiral neutrino into a single representation (for a single generation), it is customary to work with the all left-chiral formulation. The fermionic Lagrangian is written (in terms of flavor eigenstates)
\begin{equation}
\mathcal{L}_{fermion} = i q^\dagger \bar{\sigma}^\mu D_\mu q + i l^\dagger \bar{\sigma}^\mu D_\mu l + i u^\dagger \bar{\sigma}^\mu D_\mu u + i d^\dagger \bar{\sigma}^\mu D_\mu d + i e^\dagger \bar{\sigma}^\mu D_\mu e + i \nu^\dagger \bar{\sigma}^\mu D_\mu \nu).  
\end{equation}
The representations corresponding to these left-chiral fields describe the index structure of the fields for the first generation
\begin{eqnarray}
 q= q^{ia}_\alpha : (\textbf{3},\textbf{2})_{1/6} , & l=l^a_\alpha: (\textbf{1},\textbf{2})_{-1/2}, & u=u^i_\alpha: (\overline{\textbf{3}},\textbf{1})_{-2/3}, \nonumber \\
 d=d^i_\alpha: (\overline{\textbf{3}},\textbf{1})_{1/3},& e=e_\alpha: (\textbf{1},\textbf{1})_{1},& \nu=\nu_\alpha: (\textbf{1},\textbf{1})_0.  
\end{eqnarray}
where $d$ as a left-chiral state is the charge conjugated right-chiral fermion.

The interactions of the fermions are encoded in the covariant derivatives, which are given by
\begin{align}
D_\mu q &= \left( \partial_\mu  - i g_3 G_\mu  - i g_2 W_\mu - i Y_q B_\mu \right) q, \\ 
D_\mu l &= \left( \partial_\mu - i g_2 W_\mu - i Y_l B_\mu \right) l, \\ 
D_\mu u &= \left( \partial_\mu  - i g_3 G_\mu - i Y_u B_\mu \right) u, \\ 
D_\mu d &= \left( \partial_\mu  - i g_3 G_\mu - i Y_d B_\mu \right) d, \\ 
D_\mu e &= \left( \partial_\mu  - i Y_e B_\mu \right) e, \\ 
D_\mu \nu &= \partial_\mu \nu.
\end{align}
Since the spinor equations of motion contain $\gamma^\mu D_\mu \lambda$ for different spinors $\lambda$, the covariant derivatives above encode the interactions between the fermions and gauge bosons. Section ~\ref{int-ferm} focuses on testing these interactions in the helon model.


\begin{thebibliography}{100}

\bibitem{hossenfelder2013minimal}
S.~Hossenfelder.
\newblock Minimal length scale scenarios for quantum gravity.
\newblock {\em Living Reviews in Relativity}, 16(1):2, 2013.

\bibitem{kumar2020quantum}
S.~P. Kumar and M.~B. Plenio.
\newblock On quantum gravity tests with composite particles.
\newblock {\em Nature Communications}, 11(1):3900, 2020.

\bibitem{lees1939xxxvi}
A~Lees.
\newblock The electron in classical general relativity theory.
\newblock {\em The London, Edinburgh, and Dublin Philosophical Magazine and
  Journal of Science}, 28(189):385--395, 1939.

\bibitem{dirac1962extensible}
P.~A.~M. Dirac.
\newblock An extensible model of the electron.
\newblock {\em Proceedings of the Royal Society of London. Series A.
  Mathematical and Physical Sciences}, 268(1332):57--67, 1962.

\bibitem{dirac1962particles}
P.~A.~M. Dirac.
\newblock Particles of finite size in the gravitational field.
\newblock {\em Proceedings of the Royal Society of London. Series A.
  Mathematical and Physical Sciences}, 270(1342):354--356, 1962.

\bibitem{dirac1931quantised}
P.~A.~M. Dirac.
\newblock Quantised singularities in the electromagnetic field.
\newblock {\em Proceedings of the Royal Society of London. Series A, Containing
  Papers of a Mathematical and Physical Character}, 133(821):60--72, 1931.

\bibitem{Pati:1974yy}
Jogesh~C. Pati and Abdus Salam.
\newblock {Lepton Number as the Fourth Color}.
\newblock {\em Phys. Rev.}, D10:275--289, 1974.
\newblock [Erratum: Phys. Rev.D11,703(1975)].

\bibitem{Shupe:1979fv}
M.~A. Shupe.
\newblock {A Composite Model of Leptons and Quarks}.
\newblock {\em Phys. Lett. B}, 86:87--92, 1979.

\bibitem{Harari:1979gi}
Haim Harari.
\newblock {A Schematic Model of Quarks and Leptons}.
\newblock {\em Phys. Lett. B}, 86:83--86, 1979.

\bibitem{Harari:1980ez}
Haim Harari and Nathan Seiberg.
\newblock {A Dynamical Theory for the Rishon Model}.
\newblock {\em Phys. Lett. B}, 98:269--273, 1981.

\bibitem{Raitio:1979ru}
Risto Raitio.
\newblock {A Model of Lepton and Quark Structure}.
\newblock {\em Phys. Scripta}, 22:197, 1980.

\bibitem{Lehto:1981uz}
M.~Lehto and R.~Raitio.
\newblock {A Gauge Theory of Quark and Lepton Constituents}.
\newblock {\em Phys. Scripta}, 25:239, 1982.

\bibitem{Fritzsch:1981zh}
H.~Fritzsch and G.~Mandelbaum.
\newblock {Weak Interactions as Manifestations of the Substructure of Leptons
  and Quarks}.
\newblock {\em Phys. Lett. B}, 102:319--322, 1981.

\bibitem{Bars:1982zq}
Itzhak Bars.
\newblock {Theoretical and Phenomenological Constraints on Preons, Models and
  Supergroups}.
\newblock {\em Nucl. Phys. B}, 208:77--121, 1982.

\bibitem{Zenczykowski:2008xt}
Piotr Zenczykowski.
\newblock {The Harari-Shupe preon model and nonrelativistic quantum phase
  space}.
\newblock {\em Phys. Lett. B}, 660:567--572, 2008.

\bibitem{Raitio:2018ofm}
Risto Raitio.
\newblock {Supersymmetric Preons and the Standard Model}.
\newblock {\em Nucl. Phys. B}, 931:283--290, 2018.

\bibitem{Lambek:2000ek}
J.~Lambek.
\newblock {Four-vector representation of fundamental particles}.
\newblock {\em Int. J. Theor. Phys.}, 39:2253--2258, 2000.

\bibitem{Bilson-Thompson:2005mby}
Sundance~O. Bilson-Thompson.
\newblock {A Topological model of composite preons}.
\newblock 3 2005.

\bibitem{Bilson-Thompson:2006xhz}
Sundance~O. Bilson-Thompson, Fotini Markopoulou, and Lee Smolin.
\newblock {Quantum gravity and the standard model}.
\newblock {\em Class. Quant. Grav.}, 24:3975--3994, 2007.

\bibitem{Bilson-Thompson:2008cmf}
Sundance Bilson-Thompson, Jonathan Hackett, Lou Kauffman, and Lee Smolin.
\newblock {Particle Identifications from Symmetries of Braided Ribbon Network
  Invariants}.
\newblock 4 2008.

\bibitem{Bilson-Thompson:2009dvo}
Sundance Bilson-Thompson, Jonathan Hackett, and Louis~H. Kauffman.
\newblock {Particle Topology, Braids, and Braided Belts}.
\newblock {\em J. Math. Phys.}, 50:113505, 2009.

\bibitem{Bilson-Thompson:2011hue}
Sundance Bilson-Thompson, Jonathan Hackett, Louis Kauffman, and Yidun Wan.
\newblock {Emergent Braided Matter of Quantum Geometry}.
\newblock {\em SIGMA}, 8:014, 2012.

\bibitem{Bilson-Thompson:2012bvs}
Sundance Bilson-Thompson.
\newblock {Braided topology and the emergence of matter}.
\newblock {\em J. Phys. Conf. Ser.}, 360:012056, 2012.

\bibitem{coecke2011interacting}
B.~Coecke and R.~Duncan.
\newblock Interacting quantum observables: categorical algebra and
  diagrammatics.
\newblock {\em New Journal of Physics}, 13(4):043016, 2011.

\bibitem{gorard2020zx}
J.~Gorard, M.~Namuduri, and X.~D. Arsiwalla.
\newblock Zx-calculus and extended hypergraph rewriting systems i: A multiway
  approach to categorical quantum information theory.
\newblock {\em arXiv preprint arXiv:2010.02752}, 2020.

\bibitem{gorard2021zx}
J.~Gorard, M.~Namuduri, and X.~D. Arsiwalla.
\newblock Zx-calculus and extended wolfram model systems ii: fast diagrammatic
  reasoning with an application to quantum circuit simplification.
\newblock {\em arXiv preprint arXiv:2103.15820}, 2021.

\bibitem{montangero2018introduction}
S.~Montangero and E.~Montangero.
\newblock {\em Introduction to tensor network methods}.
\newblock Springer, 2018.

\bibitem{zapata2022invitation}
C.~Zapata-Carratal{\'a} and X.~D. Arsiwalla.
\newblock An invitation to higher arity science.
\newblock {\em arXiv preprint arXiv:2201.09738}, 2022.

\bibitem{zapata2023hypermatrix}
C.~Zapata-Carratal{\~a}, M.~Schich, T.~Beynon, and X.~D. Arsiwalla.
\newblock Hypermatrix algebra and irreducible arity in higher-order systems:
  Concepts and perspectives.
\newblock {\em Advances in Complex Systems (ACS)}, 26(06):1--22, 2023.

\bibitem{zapata2024diagrammatic}
C.~Zapata-Carratal{\'a}, X.~D. Arsiwalla, and T.~Beynon.
\newblock Diagrammatic calculus and generalized associativity for higher-arity
  tensor operations.
\newblock {\em Theoretical Computer Science}, 1020:114915, 2024.

\bibitem{arsiwalla2020homotopic}
X.~D. Arsiwalla.
\newblock Homotopic foundations of wolfram models.
\newblock {\em Wolfram Community}, 2020.

\bibitem{arsiwalla2021homotopies}
X.~D. Arsiwalla, J.~Gorard, and H.~Elshatlawy.
\newblock Homotopies in multiway (non-deterministic) rewriting systems as $ n
  $-fold categories.
\newblock {\em arXiv preprint arXiv:2105.10822}, 2021.

\bibitem{arsiwalla2023pregeometry}
X.~D. Arsiwalla, H.~Elshatlawy, and D.~Rickles.
\newblock Pregeometry, formal language and constructivist foundations of
  physics.
\newblock {\em arXiv preprint arXiv:2311.03973}, 2023.

\bibitem{rickles2023ruliology}
D.~Rickles, H.~Elshatlawy, and X.~D. Arsiwalla.
\newblock Ruliology: Linking computation, observers and physical law.
\newblock {\em arXiv preprint arXiv:2308.16068}, 2023.

\bibitem{arsiwalla2024pregeometric}
X.~D. Arsiwalla and J.~Gorard.
\newblock Pregeometric spaces from wolfram model rewriting systems as homotopy
  types.
\newblock {\em International Journal of Theoretical Physics}, 63(4):83, 2024.

\bibitem{arsiwalla2024operator}
X.~D. Arsiwalla, D.~Chester, and L.~H. Kauffman.
\newblock On the operator origins of classical and quantum wave functions.
\newblock {\em Quantum Studies: Mathematics and Foundations}, 11:193--215,
  2024.

\bibitem{chester2024quantization}
D.~Chester, X.~D. Arsiwalla, L.~H. Kauffman, M.~Planat, and K.~Irwin.
\newblock Quantization of a new canonical, covariant, and symplectic
  hamiltonian density.
\newblock {\em Symmetry}, 16(3):316, 2024.

\bibitem{carter1999structures}
J.~S. Carter, L.~H. Kauffman, and M.~Saito.
\newblock Structures and diagrammatics of four dimensional topological lattice
  field theories.
\newblock {\em Advances in mathematics}, 146(1):39--100, 1999.

\bibitem{aganagic2005topological}
M.~Aganagic, A.~Klemm, M.~Marino, and C.~Vafa.
\newblock The topological vertex.
\newblock {\em Communications in mathematical physics}, 254:425--478, 2005.

\bibitem{arsiwalla2006phase}
X.~D. Arsiwalla, R.~Boels, M.~Marino, and A.~Sinkovics.
\newblock Phase transitions in q-deformed 2d yang-mills theory and topological
  strings.
\newblock {\em Physical Review D}, 73(2):026005, 2006.

\bibitem{arsiwalla2008more}
X.~D. Arsiwalla.
\newblock More rings to rule them all: fragmentation, 4d $\leftrightarrow$ 5d
  and split-spectral flows.
\newblock {\em Journal of High Energy Physics}, 2008(02):066, 2008.

\bibitem{arsiwalla2009entropy}
X.~D. Arsiwalla.
\newblock Entropy functions with 5d chern-simons terms.
\newblock {\em Journal of High Energy Physics}, 2009(09):059, 2009.

\bibitem{arsiwallasupersymmetric}
X.~D. Arsiwalla.
\newblock Supersymmetric black holes as probes of quantum gravity.
\newblock {\em PhD Thesis, University of Amsterdam}, 2010.

\bibitem{salter2024ropes}
Nick Salter.
\newblock Ropes, fractions, and moduli spaces.
\newblock {\em The Mathematical Intelligencer}, pages 1--8, 2024.

\bibitem{Bern:2008qj}
Z.~Bern, J.J.M. Carrasco, and Henrik Johansson.
\newblock {New Relations for Gauge-Theory Amplitudes}.
\newblock {\em Phys.Rev.}, D78:085011, 2008.

\bibitem{Furey:2010fm}
C.~Furey.
\newblock {Unified Theory of Ideals}.
\newblock {\em Phys. Rev.}, D86:025024, 2012.

\bibitem{Furey:2016ovx}
C.~Furey.
\newblock {\em {Standard model physics from an algebra?}}
\newblock PhD thesis, Waterloo U., 2016.

\bibitem{Furey:2022qhg}
N.~Furey and M.~J. Hughes.
\newblock {One generation of standard model Weyl representations as a single
  copy of R\ensuremath{\otimes}C\ensuremath{\otimes}H\ensuremath{\otimes}O}.
\newblock {\em Phys. Lett. B}, 827:136959, 2022.

\bibitem{Ivanov:1981wm}
E.~A. Ivanov and J.~Niederle.
\newblock {Gauge Formulation of Gravitation Theories. 2. The Special Conformal
  Case}.
\newblock {\em Phys. Rev. D}, 25:988, 1982.

\bibitem{humphreys1972}
James~E. Humphreys.
\newblock {\em Introduction to Lie Algebras and Representation Theory},
  volume~9 of {\em Graduate Texts in Mathematics}.
\newblock Springer-Verlag, New York, 1972.

\bibitem{weinberg1995quantum}
Steven Weinberg.
\newblock {\em The Quantum Theory of Fields}, volume~2.
\newblock Cambridge University Press, 1995.

\end{thebibliography}

\providecommand{\noopsort}[1]{}\providecommand{\singleletter}[1]{#1}%

\end{document}